\documentclass[11pt]{amsart}

\usepackage{amsmath, amsthm, amssymb, amsfonts, enumerate, graphicx}
\usepackage[colorlinks=true,linkcolor=blue,urlcolor=blue]{hyperref}
\usepackage{dsfont}
\usepackage{color}
\usepackage{geometry}
\usepackage{eurosym}
\usepackage{graphicx}
\usepackage{subcaption}
\usepackage{todonotes}
\usepackage{subcaption}
\usepackage{cleveref}

\DeclareGraphicsExtensions{.png,.jpg,.bmp,.eps,.pdf}

\numberwithin{equation}{section}
\numberwithin{figure}{section}

\newtheorem{thm}{Theorem}[section]
\newtheorem{prop}[thm]{Proposition}

\newtheorem{remark}[thm]{Remark}

\newcommand{\cI}{\mathcal{I}}

\newcommand{\x}{\underline x}
\newcommand{\xx}{\overline x}

\newcommand{\uu}{\mathcal{U}}

\title[Optimal adoption of an electric vehicle]{Optimal switch from a fossil-fueled to an electric vehicle}

\author[Falbo, Ferrari, Rizzini, Schmeck]{Paolo Falbo, Giorgio Ferrari, Giorgio Rizzini, Maren Diane Schmeck}


\address{P.~Falbo: Dipartimento di Economia e Management, Università di Brescia, Contrada S. Chiara 50, 25122, Brescia, Italia}
\email{\href{mailto:paolo.falbo@unibs.it}{paolo.falbo@unibs.it}}
\address{G.~Ferrari: Center for Mathematical Economics (IMW), Bielefeld University, Universit\"atsstrasse 25, 33615, Bielefeld, Germany}
\email{\href{mailto:giorgio.ferrari@uni-bielefeld.de}{giorgio.ferrari@uni-bielefeld.de}}
\address{G.~Rizzini: Dipartimento di Economia e Management, Università di Brescia, Contrada S. Chiara 50, 25122, Brescia, Italia}
\email{\href{mailto:giorgio.rizzini@unibs.it}{giorgio.rizzini@unibs.it}}
\address{M.D. ~Schmeck: Center for Mathematical Economics (IMW), Bielefeld University, Universit\"atsstrasse 25, 33615, Bielefeld, Germany}
\email{\href{mailto:maren.schmeck@uni-bielefeld.de}{maren.schmeck@uni-bielefeld.de}}

\date{\today}

\numberwithin{equation}{section}

\begin{document}

\begin{abstract}
In this paper we propose and solve a real options model for the optimal adoption of an electric vehicle. A policymaker promotes the abeyance of fossil-fueled vehicles through an incentive, and the representative fossil-fueled vehicle's owner decides the time at which buying an electric vehicle, while minimizing a certain expected cost. This involves a combination of various types of costs: the stochastic opportunity cost of driving one unit distance with a traditional fossil-fueled vehicle instead of an electric one, the cost associated to traffic bans, and the net purchase cost.\ After determining the optimal switching time and the minimal cost function for a general diffusive opportunity cost, we specialize to the case of a mean-reverting process. In such a setting, we provide a model calibration on real data from Italy, and we study the dependency of the optimal switching time with respect to the model's parameters.  Moreover, we study the effect of traffic bans and incentive on the expected optimal switching time.\ 
We observe that incentive and traffic bans on fossil-fueled transport can be used as effective tools in the hand of the policymaker to encourage the adoption of electric vehicles, and hence to reduce air pollution.\ 
\end{abstract}

\maketitle

\smallskip


\smallskip

\smallskip



\section{Introduction}
\label{intro}

It is well known that air pollution is increasing each day for many reasons. The increase depends on the number of fossil-fueled vehicles on the road, industrial emissions, especially in energy sector (see \cite{Canadell}, \cite{Huising2015}), domestic heating, transportation system, and many others (cf.\ \cite{WHO2019}). Several studies have shown the connection between air pollution and human health damages, such as acute inflammation of the respiratory tract, asthma crisis, alterations in the functioning of the cardiovascular system, tumors etc.; for reference, see, for example, \cite{GautamBolia2020} and \cite{Delf}. Moreover, it is estimated that air pollution causes around 7 million premature deaths in 2016 worldwide \cite{WHO2019}.

Various studies identify road traffic emissions as one of the major contributors to air pollution in the urban area (see \cite{Colvile2001}, \cite{Belis2013}, \cite{Karagulian2015}). This is confirmed also, fo example, in \cite{COVID-ITA}, \cite{Liucovid} and \cite{COVID-CINESI}, which affirm that the traffic restriction due to the lockdown actions have decreased air pollution.\ Numerous solutions to road traffic pollution have been conjectured and, among these, fleet electrification is seen as one of the most practical.\ In \cite{Soret2014} it is shown how transportation fleet electrification could be a good solution for improving air quality with two study cases of Madrid and Barcelona cities.\ The reduction of pollutants' concentration through incentive policies for the adoption of electric vehicles is also confirmed in \cite{Zhao2017} and \cite{Laberteaux2018}. 

As many authors state (see, for example, \cite{AM} or \cite{NKR}), the road transportation electrification cannot be seen as the unique solution to the pollution problem. The road fleet electrification needs to be accompanied by an environmental policy which induces the strong use of renewable in the electricity generation. A subsidies can be a possible strategy to achieve the aim of using renewables in electricity generation, as shown in \cite{Kort2019}.  It is shown in \cite{Buekers2014} that, in terms of external costs, the economic benefit for a country adopting electric vehicles can be positive or negative depending on the country's energy generation mix. The authors show that countries with energy mix depending on renewable sources may profit in terms of avoided external costs, such as the reduction of healthcare and social costs due to air pollution. The same findings can be read in \cite{CMGN} and \cite{OKT}.

To achieve the goal of reducing emissions through vehicle fleet electrification, policymakers throughout the world face the enduring public policy problem of how to encourage the adoption of electric vehicles. Public policy actions can be divided into direct and indirect actions. The diffusion of direct and indirect actions varies across countries because of several factors that characterize each case; e.g., charging infrastructure development, electricity load distribution and management, and cultural and economical features of people (differentiated through their personal income, willingness to pay, perceived risk, as well as psychological aspects like moral values and behavior).
Direct interventions -- also called purchase-based incentive policies -- are used in order to lower the electric vehicle purchase price. They include financial incentives, reduction in registration fee, registration tax, and vehicle ownership tax. On the other hand, indirect actions -- also called use-based incentive policies -- include, for example, the implementation of infrastructures, such as charging station and support for R\&D (research and development). For the EU area, a literature review on direct and indirect actions can be found in \cite{Cansino2018}, where the different policies adopted by member states are compared.

\vspace{0.25cm}

 
\emph{Our work.} In this paper, we consider a representative owner of a fossil-fueled vehicle deciding the time at which purchasing an electric vehicle. The decision is taken with the aim of minimizing the total expected costs accrued when driving the fossil-fueled vehicle for a given distance. The first cost is the opportunity of driving one unit distance with a traditional fossil-fueled vehicle instead of an electric one, whenever fossil fuel is more expensive than electricity. The cost difference between the two fuels is assumed to be stochastic and evolving as a (one-dimensional) It\^o's diffusion. A second cost faced by the agent comes from the fact that traffic bans for fossil-fueled vehicles can happen at certain times chosen by the policymaker. We assume that those traffic stops happen at a given constant rate according to some criterion chosen by the policymaker. The introduction of the latter cost is strongly motivated by the real-world pollution's reduction policies applied, for example, in the Northern of Italy. The last cost incurred by the agent is the purchase cost of an electric vehicle, net of the possible incentive ensured by the policymaker.

With regards to the methodology, we model the agent's decision problem as an optimal stopping problem and we solve it by relying on a classical guess-and-verify approach, which is based on the construction of candidate value function and optimal stopping time whose actual optimality is then verified through a verification theorem. In order to deal with our general diffusive setting, we employ the approach of \cite{Alvarez} on optimal stopping problems for one-dimensional regular diffusions, and we show that the optimal adoption time is of barrier-type. Indeed, the fossil-fueled vehicle's owner should switch to an electric vehicle only when the the opportunity cost is sufficiently large (i.e., when the price of fossil fuel is sufficiently higher than that of electricity). 

In a case study, we take the opportunity cost evolving as an Ornstein-Uhlenbeck process. Using real data from Italy, we validate statistically the chosen dynamics of the opportunity cost process and we calibrate the model. We then apply the decision model to study the dependency of the optimal switching time with respect to the various model's parameters. Moreover, we present the study of the expected switching time with respect to frequency of traffic bans and incentive.\ In particular, we show that economic incentives and the frequency of traffic bans on fossil-fueled transport can be effective tools in the hand of the policymaker to encourage the adoption of electric vehicles, and hence to reduce pollution.\ 
\vspace{0.25cm}

\emph{Related literature.} Several papers consider the decision of switching to an electric vehicle for a representative consumer/vehicle owner using a real options approach. In \cite{MoonLee2019} it is investigated the decision of a consumer that aims at purchasing an electric vehicle by taking into account the total cost of ownership. The latter depends on fuel price uncertainty (modeled through a discrete binomial model) and technological advancements. Data are taken from the Korean market and the authors illustrate that, even without incentives, electric vehicles are more cost-effective than fossil-fueled ones. Moreover, the authors show that, when the uncertainty in fuel price decreases, people willingness to purchase an electric vehicle increases. \cite{Agaton2019} presents an investment problem for a transportation operator deciding whether to purchase a fossil-fueled jeeps or electric jeeps fleet, by considering a governmental incentive and diesel price uncertainty. The latter follows geometric Brownian dynamics. The authors show that, considering the current Philippines' market structure, the optimal decision is to purchase a fleet of electric jeeps immediately. Moreover, it is suggested to governments to increase the incentive, the implementation of public charging stations, and the use renewable sources for electricity generation. A decision problem for a consumer who minimizes her transport expenditures is addressed in \cite{HeFanLiLi2017}. The representative consumer wishes to optimally replace her old fossil-fueled vehicle by an hybrid electric vehicle. To this purpose, she considers government's incentives for the replacement and fossil fuel price uncertainty (modeled as a geometric Brownian motion). The authors show that incentives, uncertainty in fossil fuel price and distance traveled affect the replacement time. Moreover, when fuel price increases, the effect of incentive is attenuated, while a decrease in fuel price volatility reduces the replacement time. 

Among the works taking instead the point of view of a company producing cars or of a policymaker, we refer to \cite{Nishihara2010}, \cite{KangBayarakPapalambros2018}, \cite{Yamashita}, and \cite{Ansaripoor}. In \cite{Nishihara2010} the authors consider an automaker's whose cash-flows are modeled through a geometric Brownian motion. The automaker aims at promoting an hybrid vehicle, with the option to change the promotion to an electric vehicle in a future time. \cite{KangBayarakPapalambros2018} investigates the decision of a manufacturer deciding the optimal time at which launch a new vehicle segment or to redesign the existing one under market uncertainty, which is modeled through gas price and regulatory standards. The decisions refer to the optimal timing and the choice of vehicles engine-type, which can be gas, electric and hybrid. In \cite{Yamashita} it is investigated the development of plug-in electric vehicles' market by considering fuel price uncertainty and the availability of charging infrastructure. The policymaker has to determine the optimal incentive policy that can be used either to reduce the electric vehicle net purchase price or to build new charging infrastructures. The authors show that an increase in fuel price leads to a decrease in the incentive's effectiveness and to an increase in the electric vehicles' adoption. Finally, \cite{Ansaripoor} studies the use of flexible lease contracts to determine the optimal number of vehicles to be leased minimizing simultaneously the risk (Recursive Expected Conditional Value at Risk) and the costs. The firm has to decide which type of engine among fossil-fueled, hybrids, and electric has to be used. Uncertainty is about CO$_2$ prices, fuel prices, distance traveled, fuel consumption, and technological aspects. The authors show that, when considering the technological change of electric vehicles, electric vehicles are the preferred technology for leasing.\\

Our work contributes to the literature employing a real options approach for the adoption of electric vehicles. In particular, differently to the existing works, we are able to accommodate in our model the possibly general and mean-reverting behavior of the opportunity cost, as well as the costs incurred by the representative agent because of the environmental policies employed by the policymaker in the form of traffic bans for fossil-fueled vehicles. To the best of our knowledge, the presence of these modeling features appear here for the first time within the literature on electric vehicles' optimal adoption problem.
\vspace{0.25cm}

The rest of the paper is organized as follows. Section \ref{problemsetting} formulates the problem within a general diffusive setting and presents its solution. Section \ref{Model Analysis} is devoted to the case study in which the opportunity cost process follows a mean-reverting dynamics. Section \ref{numerics} provides parameters' calibration, theoretical and numerical sensitivity results. Section \ref{Opt_time} provides a numerical simulation of the expected optimal switching time. Finally, Section \ref{Conclusion} collects concluding remarks on policy implications, while Appendix \ref{AppedixA} recaps some mathematical details related to one-dimensional regular diffusions and Appendix \ref{AppendixB} presents the proof the main theorem.


\section{Problem formulation and Solution}
\label{problemsetting}

\subsection{Problem Formulation}
\label{problemformulation}

We consider a fossil-fueled vehicle owner that aims, over an infinite time horizon, to manage her mobility costs. Buying an electric vehicle, she receives a fixed amount of money by a policymaker that wants to boost the adoption of green transport in order to reduce air pollutants' concentration. The fossil-fueled vehicle owner drives for an average distance of $\ell$ (per unit of time) and her aim is to determine the time at which to adopt an electric vehicle, while minimizing a certain cost functional.
The latter is composed by different types of costs. The first involves the running stochastic opportunity cost per unit distance $\{X_t\}_{t\geq0}$ of driving a fossil-fueled vehicle instead of an electric one. The second is the marginal cost $c$ associated to each traffic ban for fossil-fueled vehicles occurring at some random time. The policymaker, for environmental reasons, imposes traffic bans at certain exogenous (random) times.\footnote{In many Italian cities, air pollution has spiked above the safety threshold of $50$ $\frac{\mu g}{m^3}$ for many consecutive days and several cities have introduced restrictions on driving such as a ban on fossil-fueled vehicles. Air pollution is typically worst in Northern Italy, where densely populated cities, industry and farming create emissions and mountains trap it in low-lying plains. In the Italian context, we refer to the \emph{Temporary Limitations of 1}$^{\text{\emph{st}}}$\emph{Level to Road Traffic} of the document \textit{New Program Agreement for the Coordinated and Joint Adoption of Measures to Improve Air Quality in the Po Valley} signed in Bologna on June $9^{\text{th}}$, $2017$ by Italian Minister Galletti and the presidents of the regions of Po Basin (Emilia Romagna, Veneto, Lombardy and Piedmont) and adopted by Lombardy region with D.G.R.\ Number $X/6675$ of July $06^{\text{th}}$, $2017$. See \cite{POVALLEY}.}
The third is the purchase cost of the electric vehicle $I$, net of the financial incentive, $k$, ensured by the policymaker. We assume, as natural, that $I>k>0$. 

With this specification, the total expected cost incurred up to the switching time $\tau$ is 
\begin{equation}
\mathbb{E}\bigg[  \int_{0}^{\tau} e^{-\rho t} \ell X_{t}dt+\int_{0}^{\tau} e^{-\rho t} c dN_{t}+ e^{-\rho \tau} \left(  I-k\right) \bigg].
\label{primaformula}
\end{equation}
Here, $\rho>0$ is a personal discount factor and $\{N_t\}_{t\geq0}$ is a counting process of the traffic bans. The expectation is taken with respect to the joint law of the process $X$ and $N$, defined on some common complete filtered probability space $(\Omega, \mathcal{F}, \mathbb{F}:=(\mathcal{F}_t)_t, \mathbb{P})$. Moreover, we assume that no anticipation is allowed and that any decision is taken with respect to (w.r.t.)\ the flow of information $\mathbb{F}$; that is, we let $\tau$ be an $\mathbb{F}$-stopping time.


Assuming that the process $\left\{N_{t}\right\}  _{t\geq0}$ is an homogeneous Poisson process with constant intensity $\lambda$, independent from $\left\{X_{t}\right\}  _{t\geq0}$, and noting that the compensated Poisson process $\left\{  N_{t}-\lambda t\right\}  _{t\geq0}$ is a martingale, by the Optional Sampling Theorem (see, e.g., Theorem 3.22 in Chapter $1$ of \cite{KS}), we have (up to a standard localization procedure) 
\begin{equation*}
\mathbb{E}\bigg[\int_{0}^{\tau}c e^{-\rho t} dN_{t}\bigg] =\mathbb{E}\bigg[\int_{0}^{\tau} c e^{-\rho t}\lambda dt\bigg]
\text{.} 
\label{Poisson}
\end{equation*}
Hence, the cost functional \eqref{primaformula} equivalently rewrites 
\begin{equation}
\mathbb{E} \left[  \left(  \int%
_{0}^{\tau}e^{-\rho t}  \ell X_{t}dt+\int_{0}^{\tau}e^{-\rho t} \lambda c dt\right)  +e^{-\rho\tau} \left(  I-k\right)  \right]  \text{,}%
\label{formaequivalente}
\end{equation}
where we see that the fossil-fueled vehicle owner faces now a running cost $\lambda c$ associated to an average number $\lambda$ of traffic bans per unit of time. Notice that in \eqref{formaequivalente} we do not consider any vehicle maintenance costs since those must be paid both for an electric and a fossil-fueled vehicle, and they can be supposed of similar value.

Then, the fossil-fueled vehicle owner aims at finding when to switch to an electric vehicle so that her total costs are minimized; that is, she aims at solving
\begin{equation}
\inf_{\tau\geq0}\mathbb{E}\left[  \int_{0}^{\tau} e^{-\rho t} \ell X_{t}dt+\int_{0}^{\tau}e^{-\rho t} \lambda c dt+ e^{-\rho\tau}\left(  I-k\right)  \right]  \text{.} \label{Funzionale}
\end{equation}


From \eqref{Funzionale} we see that the decision of the fossil-fueled vehicle owner should only depend on the evolution of the process $\left\{X_t \right\}_{t\ge0}$; that is, on the opportunity cost of driving one unit distance with a fossil-fueled vehicle instead of an electric one (in the sequel, we shall also write ``the unit distance opportunity cost''). 

The process $\left\{X_t \right\}_{t\ge0}$ is assumed to evolve according to the stochastic differential equation (SDE)
\begin{equation}
dX_t =\mu (X_t)dt+ \sigma(X_t)dW_t,
\label{SDE}
\end{equation}
with intial condition $X_0=x \in \cI$, $\cI :=(\underline{x}, \overline{x})$, with $- \infty \le \underline{x} < \overline{x} \le +\infty$. Here, $\mu, \sigma: \cI \to \mathbb{R}$ are Borel-measurable and such that
\begin{equation}
\label{non_degeneracy}
\sigma^2 \left( x \right) >0, \quad x \in \cI\text{,}
\end{equation}
and, $\forall x \in \cI$, there exists $\varepsilon>0$ (depending on $x$) such that
\begin{equation}
\label{local_integrability}
\int_{x - \varepsilon}^{x +\varepsilon} \frac{1+ | \mu \left(z \right) |}{\sigma^2 \left(z \right)} dz < \infty.
\end{equation}

\begin{remark}
\eqref{non_degeneracy} expresses a nondegeneracy condition, while \eqref{local_integrability} is a local integrability condition.
\end{remark}


According to the results in Chapter $5$ of \cite{KS}, SDE \eqref{SDE} admits a weak solution $( \Omega, \mathcal{F}, \mathbb{F}, \mathbb{P}_x,$  $W,X )$. Moreover, the stochastic basis is unique in the sense of probability law (see, e.g., \cite{KS}, Theorem $5.15$ in Chapter V) and from now on, we will consider the stochastic basis $\left( \Omega, \mathcal{F}, \mathbb{F}, \mathbb{P}_x, W,X \right)$ given and fixed. Moreover,  in order to stress the dependency of $X$ on its initial level $x$, we denote the solution associated to (\ref{SDE}) as $\left\{X^x_t\right\}_{t\ge0}$. Finally, in the rest of the paper, we use the notation
$\mathbb{E}\left[f\left(  X_{t}^{x}\right)\right]  =\mathbb{E}_{x}\left[f\left(  X_{t}\right)  \right]$, where $\mathbb{E}_{x}$ represents the expectation conditioned on $X_0=x$.

Under the previous assumptions on $\mu$ and $\sigma$, the process $\left\{X_t^x\right\}_{t\ge0}$ is a regular diffusion; that is, starting from $x$, it can reach any other $y \in \cI$ in finite time with positive probability. In light with our subsequent application, we also assume that the boundary points $\underline{x}$ and $\overline{x}$ of the domain $\cI$ are natural for $X^x$; that is, they cannot be reached in finite time with positive probability (see Chapter I of \cite{BS} for further details). 

\begin{remark}
The assumptions on the process $X$ made so far are satisfied in relevant cases, as in the case of Brownian motion with drift (i.e., $\mu\left(x\right)  =\mu \in \mathbb{R}$ and $\sigma\left(  x\right)  =\sigma>0$), and in the case of Ornstein-Uhlenbeck process (i.e., $\mu\left(  x\right)  =\theta\left(
\mu-x\right)  $, for some constants $\theta>0$, $\mu\in\mathbb{R}$ and $\sigma\left(  x\right)  =\sigma>0$). Those processes take values on $\mathcal{I}=\mathbb{R}$, so that $\underline{x}=-\infty$ and $\overline{x}=+\infty$.
\end{remark}

Since the value of the stopping functional \eqref{formaequivalente} only depends on the law of the unit distance opportunity cost process $X$, given the underlying Markovian structure, we can define the value of the problem as
\begin{equation}
\label{Funzionale_finale}
V\left(  x\right)  :=\inf_{\tau\geq0}\mathbb{E}_{x}\left[  \int_{0}^{\tau
} e^{-\rho t} \left(  \ell X_{t}+\lambda c \right)  dt+ e^{-\rho\tau} \left(I-k\right)
\right], \quad x \in \mathcal{I},
\end{equation}
where $x$ represents the initial condition of process $X$, and the optimization is taken over the set of $\mathbb{F}$-stopping times.

The expectation in \eqref{Funzionale_finale} can be clearly rewritten as
\begin{equation*}
V\left(  x\right)  =\mathbb{E}_{x}\left[  \int_{0}^{\infty} e^{-\rho t}  \left(
\ell X_{t}+\lambda c \right)  dt\right]
+\inf_{\tau\geq
0}\mathbb{E}_{x}\left[ - \int_{\tau}^{\infty} e^{-\rho t} \left(  \ell X_{t}+\lambda c \right)  dt+e^{-\rho\tau} \left(  I-k\right) \right]
\text{,}
\end{equation*}
where the first expectation above is the value of the option of never switching to an electric vehicle. Then, setting
\[
\mathcal{U}\left(  x\right)  := \inf_{\tau \ge 0} \mathbb{E}_{x}\left[ - \int_{\tau}^{\infty} e^{-\rho t} \left(  \ell X_{t}+\lambda c\right)  dt+e^{-\rho\tau} \left(  I-k\right) \right]
\]
and
\[
 \widehat{V}\left( x \right):= \mathbb{E}_{x}\left[  \int_{0}^{\infty} e^{-\rho t} \left(
\ell X_{t}+\lambda c\right) dt\right] ,
\]
we have that
\begin{equation}
V\left(  x\right)  =\widehat{V}\left(  x\right)  +\mathcal{U}%
\left(  x\right)  \text{,} \label{equazioneValore}%
\end{equation}
where, by the help of strong Markov property,
\begin{equation}
\mathcal{U}\left(  x\right)  =\inf_{\tau\geq0}\mathbb{E}_{x}\left[
e^{-\rho\tau}\left(  \left(  I-k\right)  -\widehat{V}\left(  X_{\tau}\right)
\right)  \right]  \text{.} \label{U}%
\end{equation}

In the rest of this paper we make the standing assumption that 
\begin{equation}
\mathbb{E}_x \left[ \int_0^{\infty} e^{- \rho t} |X_t| dt \right] < \infty
\label{abs}
\end{equation}
(so that $|V|<\infty$), and for any Borel-measurable function $g$ we adopt the convention
\begin{equation}
e^{-\rho \tau} g\left( X^x_{\tau}\right) := \varliminf_{t \uparrow \infty} e^{-\rho t} g\left( X_t^x\right) \hspace{2mm} \text{on} \hspace{2mm} \{\tau = \infty\}.
\label{make_sense}
\end{equation}

The latter, in particular, ensures that the stopping cost $e^{-\rho\tau}(I-k -\widehat{V}(X_{\tau}^{x}))$ is well-defined on the event $\{\tau = \infty\}$.


\begin{remark}
Notice that \eqref{abs} has to be verified as a case by case basis. For example, it always holds true if $X$ is a Brownian motion or an Ornstein-Uhlenbeck process. It is satisfied when $\rho > \mu$ if $X$ is a geometric Brownian motion.
\end{remark}


From now on we focus on the optimal stopping problem \eqref{U} since it shares the same optimal strategy with $V$ (cf.\ \eqref{Funzionale_finale}), and since we can then obtain $V$ from \eqref{equazioneValore}.


\subsection{Problem solution}
\label{Heuristic discussion and solution}

We can reasonably expect that the agent would adopt an electric vehicle once the opportunity cost is sufficiently large. Hence, we can conjecture that there exists some critical level $x^* \in \overline{\mathbb{R}}$  such that $\tau^*=\inf\{t\ge0: X_t^x \ge x^*\}$ is optimal for \eqref{U} (hence for $V$). The following theorem provides the complete description of the problem's solution within the general setting described in the previous section. Its proof can be found in Appendix \ref{AppendixB}, where we follow the approach of \cite{Alvarez}.

\begin{thm}
\label{theorem_solution}
Let $\mathcal{L}_X$ be the infinitesimal generator of the process $X$ given by the second-order linear differential operator
\[
\mathcal{L}_X = \frac{1}{2} \sigma^2 \left( x \right) \frac{d^2}{dx^2} + \mu \left( x \right) \frac{d}{dx} \text{,}
\]
acting on two-times continuously differentiable functions. Let $\psi_{\rho}(x)$ be the strictly increasing fundamental solution to $ \left(\mathcal{L}_{X} u-\rho u\right)=0$ and $m'$ be the speed measure density of $X$ (cf.\ Appendix \ref{AppedixA} for details). We then have the next three cases.
\vspace{0.25cm}

\emph{(a)}
If
\begin{equation}
\lim_{x \to \x} \left( \ell x + \lambda c \right) < \rho \left( I-k\right) <\lim_{x \to \xx} \left( \ell x + \lambda c \right)
\label{Astar}
\end{equation}
and
\begin{equation}
\lim_{x \to \x}   \frac{   \psi_{\rho}\left(  x \right) \left(  I-k-\widehat{V}\right)  ^{\prime
}\left(  x \right)  -   \psi_{\rho}^{\prime}\left(  x \right) \left(  I-k-\widehat{V} \right)\left( x \right) }{S^{\prime} \left(x \right) } =0,
\label{Axbar}
\end{equation}
then one has
\begin{equation}
\mathcal{U}\left(  x\right)  =\left\{
\begin{array}
[c]{lll}%
\left(  I-k-\widehat{V}\left(  x^* \right)  \right)  \frac{\psi_{\rho
}\left(  x\right)  }{\psi_{\rho}\left(  x^{\ast}\right)  }\text{,} &
\text{for} & x<x^{\ast}  \text{,}\\
I-k-\widehat{V}\left(  x\right)  \text{,} & \text{for} & x\geq x^{\ast}  \text{,}%
\end{array}
\right.
\label{ucapp}
\end{equation}
where $x^*$ is the unique solution in $\left( \hat{x},\infty\right)$, $\hat{x}:= \frac{1}{\ell} \left(\rho \left( I-k\right)-\lambda c\right)$, to the integral equation
\begin{equation}
\label{eq:int-eq}
\int_{\underline{x}}^{x^{\ast}}\psi_{\rho}\left(
y\right)  m^{\prime}(y) \big(\ell y + \lambda c -  \rho\left(  I-k\right)\big)  dy=0.
\end{equation}
Moreover $\tau^* := \inf\{t\ge0: X_t^x \ge x^*\} $ is optimal for $\uu$ (hence for $V$).
\vspace{0.25cm}

\emph{(b)} If $\lim_{x \to \xx} \left( \ell x + \lambda c \right) < \rho \left( I-k \right) $, then it is never optimal to switch; i.e. $\widehat{\mathcal{U}}\left(x \right) =0$, $x \in \cI$, and $\tau^* = +\infty$ $\mathbb{P}_x$-a.s.\ (hence $V\left( x \right) = \widehat{V} \left( x \right)$).
\vspace{0.25cm}

\emph{(c)}If $\lim_{x \to \x} \left( \ell x + \lambda c \right) > \rho \left( I-k \right) $, then it is optimal to switch immediately; i.e. $\widehat{\mathcal{U}}\left(x \right) = I-k - \widehat{V} \left(x \right)$, $x \in \cI$, and $\tau^* =0$ $\mathbb{P}_x$-a.s.\ (hence $V\left(x \right) = I-k$).
\end{thm}


\section{Model's Analysis in a case study with mean-reverting dynamics}
\label{Model Analysis}

In this section we assume that in \eqref{SDE}, one has $\mu(x) = a-bx $ for some constants $b>0$, $a\in\mathbb{R}$ and $\sigma(x)  =\sigma>0$. This means that the unit distance opportunity cost is described by a mean-reverting process with dynamics
\begin{equation}
dX_{t}^{x}=\left(  a-bX_{t}^{x}\right)  dt+\sigma dW_{t}\text{, \ }X_{0}%
^{x}=x\in\mathbb{R}\text{.} \label{SDE_OU}%
\end{equation}

Here $\frac{a}{b}$ represents the mean-reversion level and $b$ is the mean-reversion speed.
The characteristic Equation \eqref{ODE_Diff} associated to \eqref{SDE_OU} is
\begin{equation}
\frac{1}{2}\sigma^{2}v^{\prime\prime}\left(  x\right)   +   \left(  a-bx\right)  v^{\prime}\left(  x\right)- \rho v\left(  x\right)=0\text{,}
\label{Lu}
\end{equation}
whose strictly increasing fundamental solutions is (cf.\ page 280 in \cite{Jeanblanc}) is
\begin{equation}
\label{psi-OU}
\psi_{\rho} \left(  x\right)  =e^{\frac{\left(  bx-a\right)  ^{2}}{2\sigma^{2}b}%
}D_{-\frac{\rho}{b}}\left(  -\frac{\left(  bx-a\right)  }{\sigma b}\sqrt
{2b}\right)  \text{,}%
\end{equation}
where $D_{\theta}\left( \cdot \right)$ is the Cylinder function of order $\theta$ (Chapter VIII in \cite{BS}) defined as
\begin{equation}
\label{cylinder-fc}
D_{\theta}\left( y\right):= e^{-\frac{x^{4}}{4}}\frac{1}{\Gamma\left(
-\theta \right)  }\int_{0}^{\infty}t^{-\theta-1}e^{-\frac{t^{2}}{2}-yt}dt, \hspace{2mm} \theta<0,
\end{equation}
and $\Gamma\left(\cdot \right)$ is the Euler's Gamma function.

Under the dynamics \eqref{SDE_OU}, 
$$\widehat{V} \left(x \right) = \frac{a}{b} \frac{\ell}{\rho} + \left( x- \frac{a}{b}\right) \frac{\ell}{ \rho + b} + \frac{ \lambda c}{\rho}.$$
Moreover, the trigger level $x^*$ can be completely characterized through the algebraic equation\footnote{As a matter of fact, \eqref{Statica} is equivalent to the integral equation \eqref{eq:int-eq} -- see \eqref{24} and \eqref{x_ottimo} in Appendix \ref{AppendixB} -- and can be obtained via imposing the classical ``smooth-pasting'' and ``smooth-fit'' conditions.}
\begin{equation}
\label{Statica}
A^{OU}(x^{\ast}) =0, 
\end{equation}
where
$$A^{OU}(x):=\left(x-\beta \right)\psi_{\rho}^{\prime}(x)  -\psi_{\rho}(x), \quad x \in \mathcal{I},$$
and
\begin{equation}
\label{beta}
\beta:=\left(  I-k-\frac{a}{b}\frac{\ell }{\rho}-\frac{\lambda c}{\rho
}\right)  \frac{\rho+b}{\ell }+\frac{a}{b}.
\end{equation}
Moreover, in this case $\underline{x}=-\infty$ and $\overline{x}=+\infty$, so that \eqref{Astar} holds. Finally, it can also be checked that \eqref{Axbar} is satisfied.

\section{An application of the decision model: a Mean-Reverting case study}
\label{numerics}
\subsection{Parameters estimation} 
\begin{figure}
\begin{subfigure}[b]{0.475\textwidth}
           \includegraphics[scale=0.5]{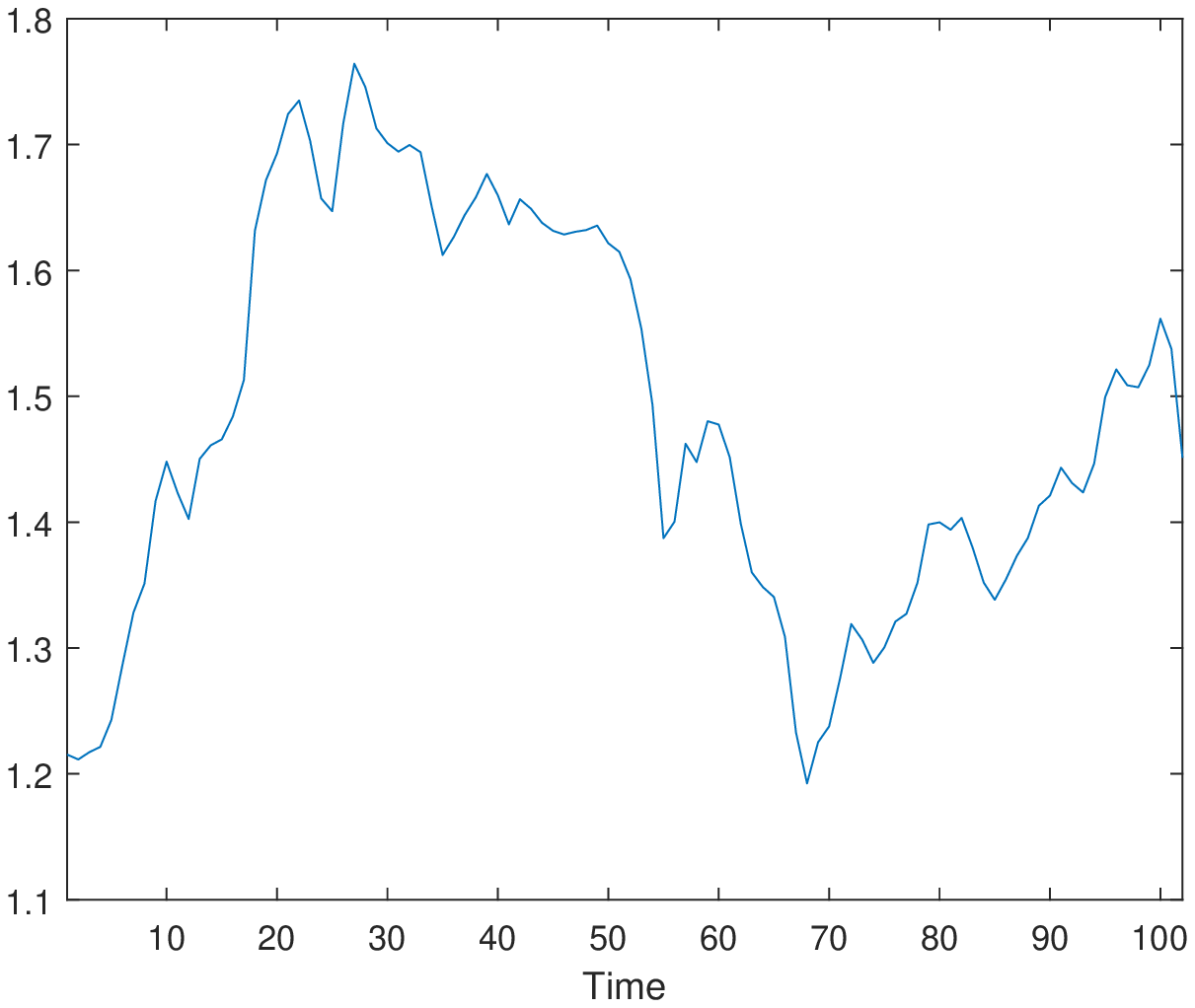}
\label{fig: plotfuel}
 \caption[]%
    {{\small Fuel price time series, $p^f_t$.}}
        \end{subfigure}
\hfill
\begin{subfigure}[b]{0.475\textwidth}
           \includegraphics[scale=0.5]{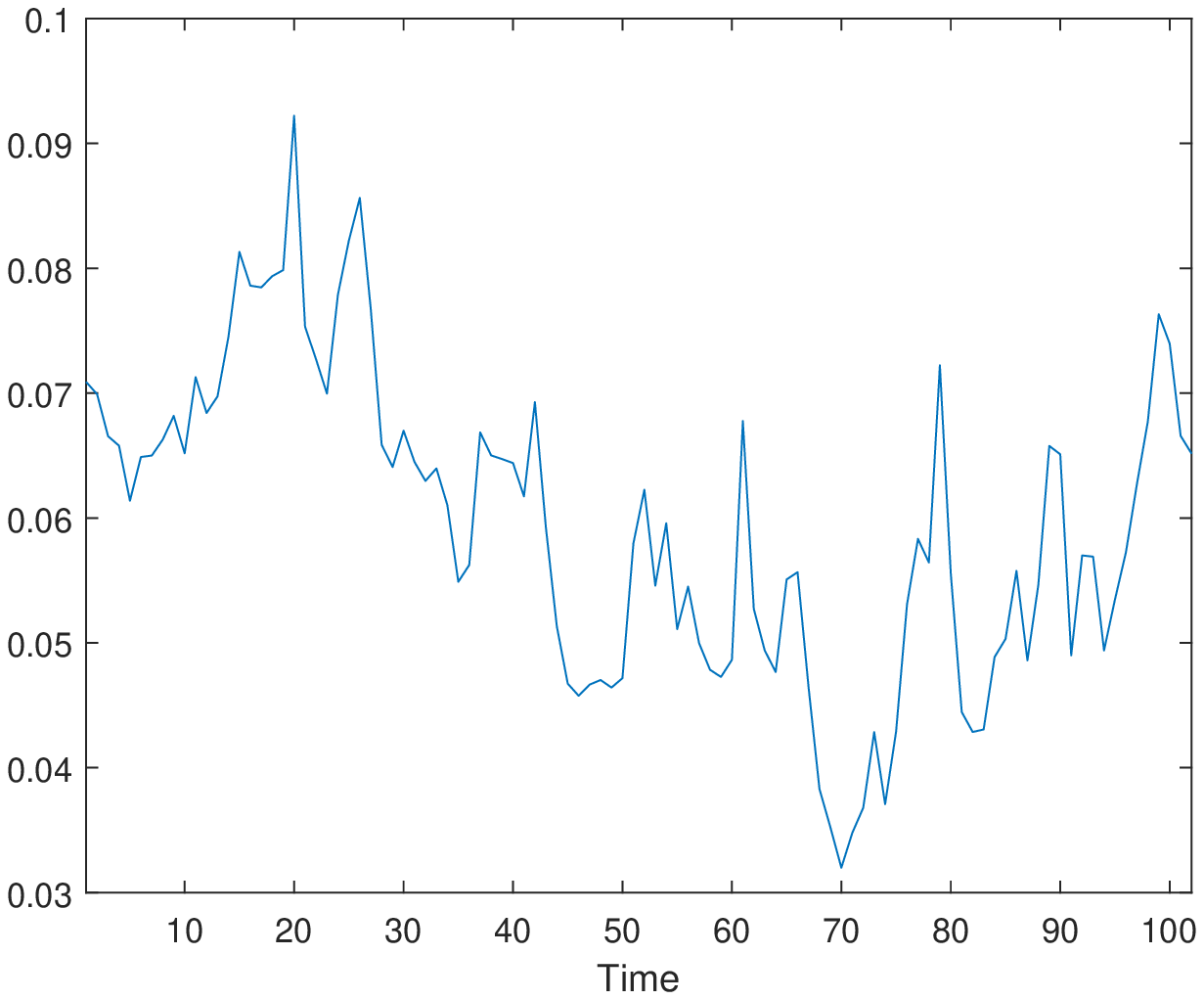}
\label{fig: plotelectricity}
 \caption[]%
    {{\small Electricity price time series, $p^e_t$.}}
        \end{subfigure}
\caption[St]
        {\small Fuel and electricity prices time series. The data are considered with monthly frequency.}
        \label{fig:fuel_electricity_plot}
\end{figure}

We now provide a calibration on Italian data of the previous mean-reverting model. We obtain the time series of the unit distance opportunity cost per kilometer $X$ as
\begin{equation}
X_t = p^f_t \cdot h^f - p^e_t \cdot h^e,
\label{time_series}
\end{equation}
where $p^f_t$ represents the  fuel price at time $t$, $h^f$ is the fuel economy of a fossil-fueled vehicle, $p^e_t$ represents the electricity price at time $t$ and $h^e$ is the fuel economy of an electric vehicle. The time series of $p^f_t$ and $p^f_t$ are plotted in Figure \ref{fig:fuel_electricity_plot}.

\begin{figure}
\includegraphics[scale=1]{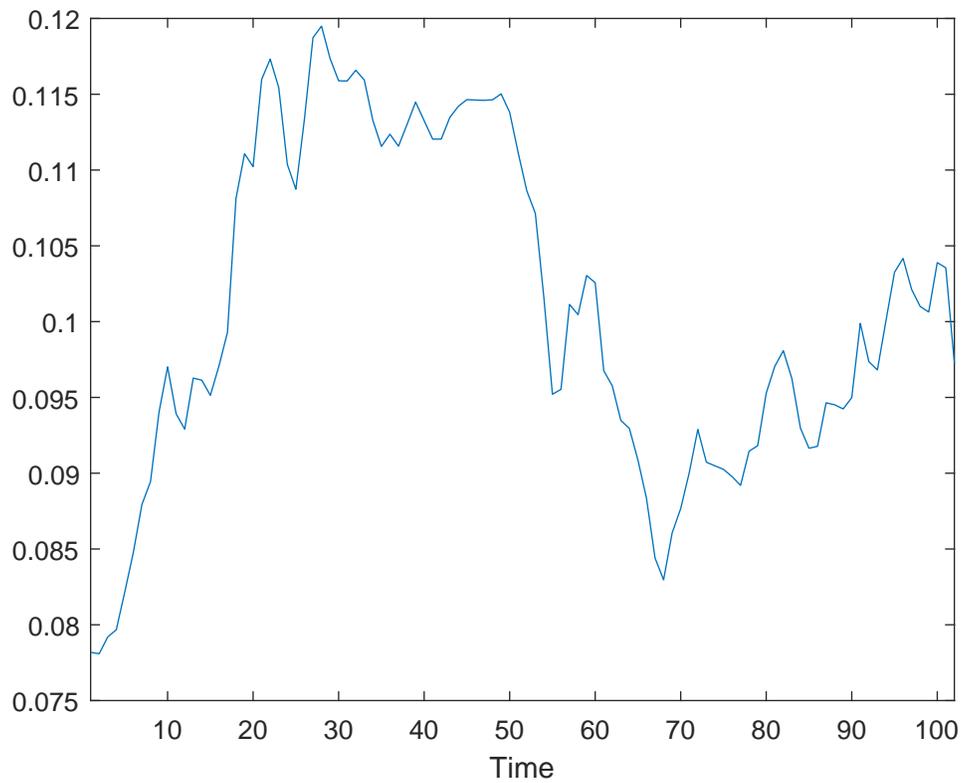}
\caption{Unit distance opportunity cost time series. The data are considered with monthly frequency.}
\label{fig: plotX}
\end{figure}

The data for $p_t^f$ and $p_t^e$ refer to Italian markets considering monthly data form July $2010$ to December $2018$, published in \cite{GOV} and \cite{TERNA}. The average monthly gasoline price, $p^f_t$, is supplied by \cite{GOV} and the average daily electricity price is supplied by \cite{TERNA}. The average monthly electricity price, $p^e_t$, is calculated averaging the daily electricity prices.
We set the fuel consumption of a fossil-fueled vehicle, $h^f$, at $0.076$ liter/km and the energy consumption of an electric vehicle, $h^e$, at $0.20$ kWh/km. Both parameters are taken as an average of technical values proposed by \cite{Plotz2012} for different sizes of fossil-fueled and electric vehicles. The unit measure of $X_t$ is euro per kilometer and, therefore, the distance is measured in kilometers. From Figure \ref{fig: plotX}, one can observe a positive correlation between the fuel and electricity price time series. Indeed, Figure \ref{fig: plotX} shows a time trend of order $10^{-5}$. We test the latter to be negligible. The hypothesis test is confirmed with a $p$-value of $0.2970$. Annual parameters estimations are made by applying the Ordinary Least Squares method to an auto-regressive model of order one. The estimated mean-reversion level of $X$, $\frac{a}{b}$, is $ 0.1045$, the estimated mean-reversion speed of $X$, $b$, is $0.5941$ and the estimated volatility parameter of $X$, $\sigma$, is $ 0.090$.

\begin{figure}
 \begin{subfigure}[b]{0.475\textwidth}
            \centering
           \includegraphics[scale=0.5]{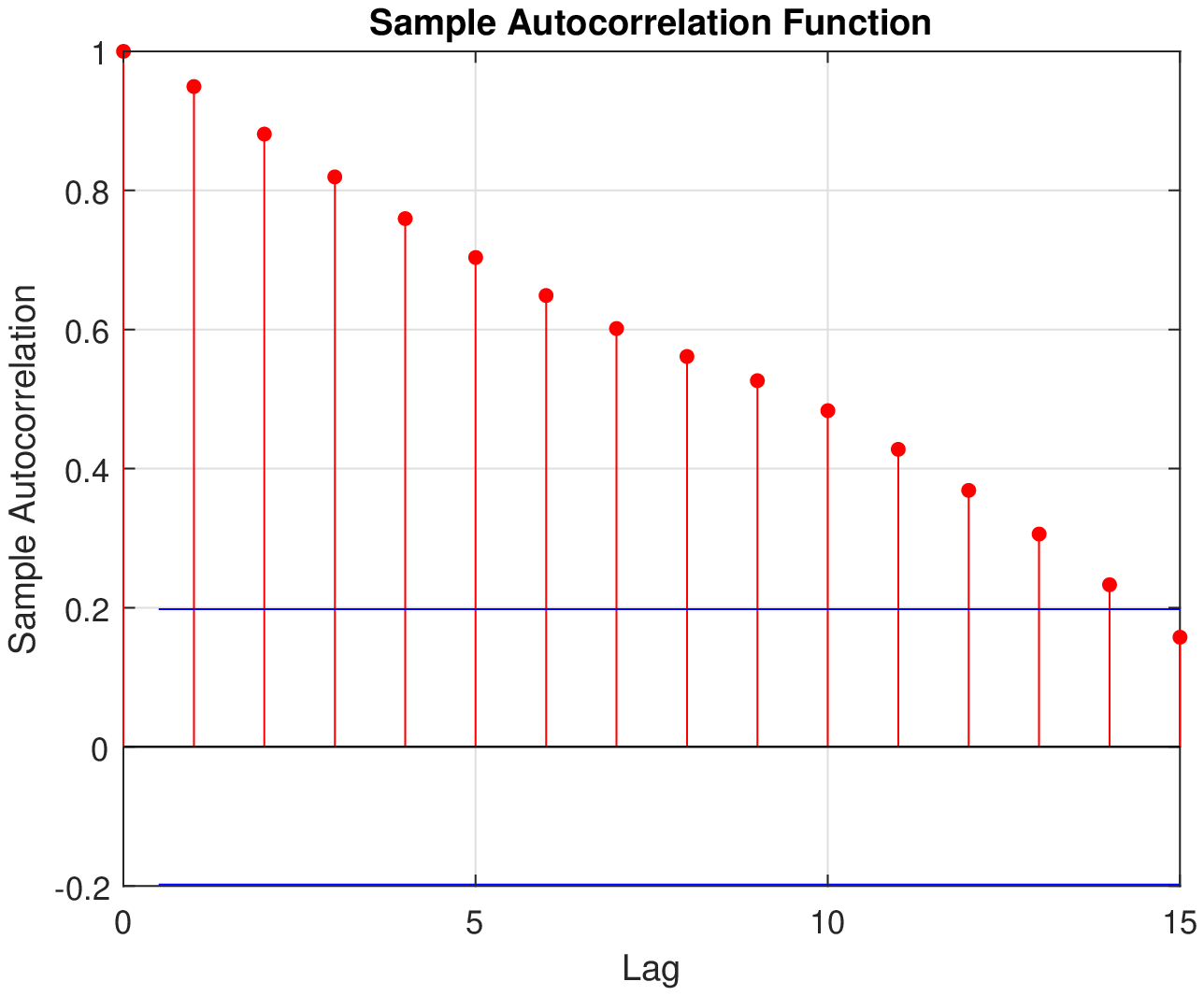}
\label{fig: ACF_X}
 \caption[]%
    {{\small Representation of the autocorrelation function for $X_t$.}}
        \end{subfigure}
        \hfill
\begin{subfigure}[b]{0.475\textwidth}
           \includegraphics[scale=0.5]{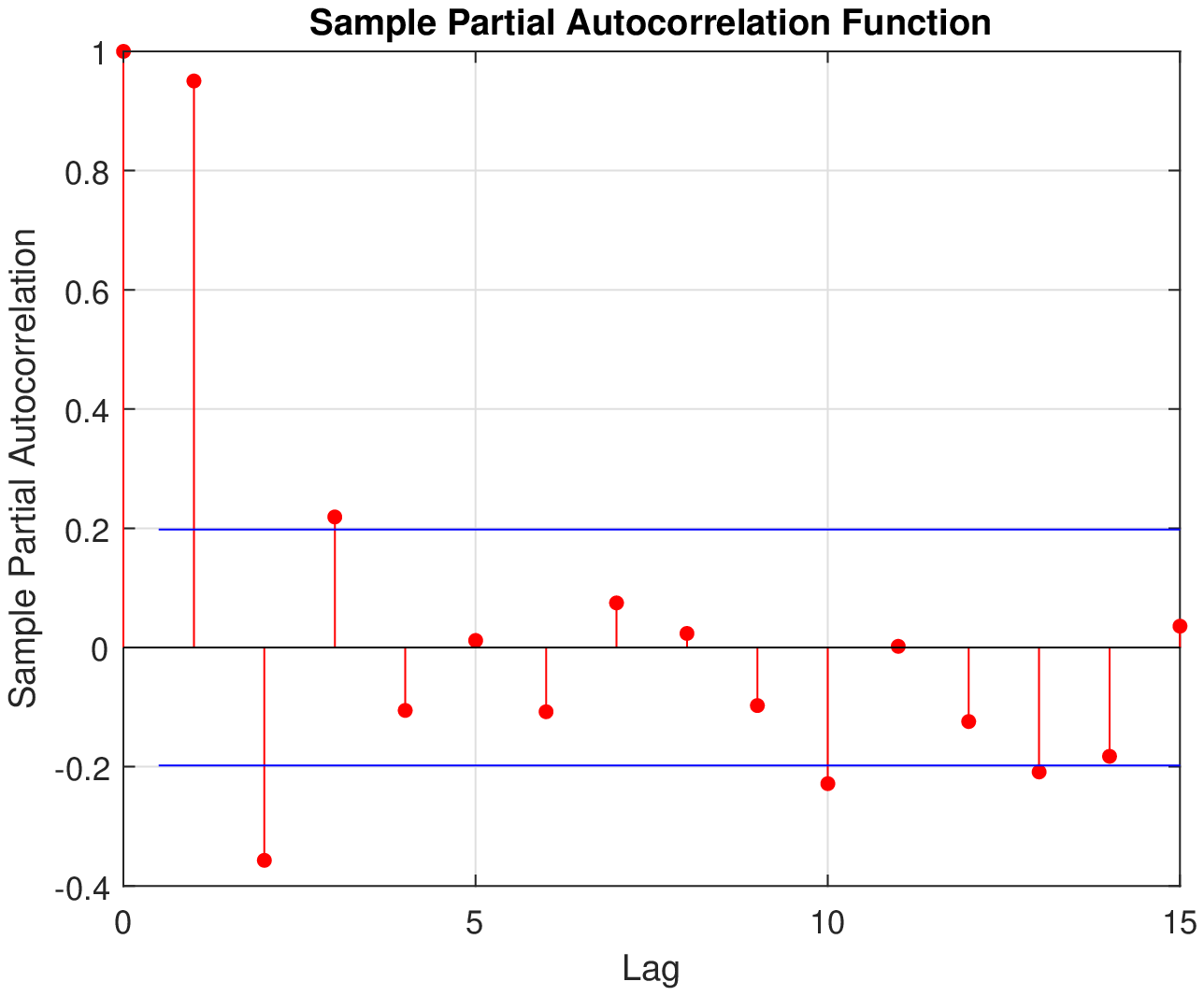}
\label{fig: PACF_X}
 \caption[]%
    {{\small Representation of the partial autocorrelation function for $X_t$.}}
        \end{subfigure}
\caption[St]
        {\small Representation of the autocorrelation and partial autocorrelation function for $X_t$. The data are considered with monthly frequency.}
        \label{fig: ACF_PACF_X}
\end{figure}

The time series of $X_t$ is plotted in Figure \ref{fig: plotX}. The graphs of auto-correlation and partial auto-correlation functions of $X_t$ are shown in Figure \ref{fig: ACF_PACF_X}. From the previous figures, we deduce that empirical data confirm the mean-reverting behavior of $X_t$ assumed in \eqref{SDE_OU}.

Based on \cite{AGENS}, the fossil-fueled vehicle owner drives an average annual distance of $12000$ kilometers. As explained in Section \ref{problemformulation}, the policymaker can decide to impose traffic bans to temporarily decrease the excess of pollutants in the air. In the Northern of Italy, for example, a traffic ban happens if the concentration of Particulate Matter $10$, PM$_{10}$, exceeds a safety threshold of $50$ $\frac{\mu g}{m^3}$ for $4$ consecutive days \cite{POVALLEY}. In this work, we study the effect of the incentive on the electric vehicle adoption and we therefore assume that the policymaker imposes traffic bans at some random times which is described by a Poisson process. The intensity $\lambda$, which represents the expected number of traffic bans, has been estimated from the regions of the Northern of Italy. The decision of imposing a traffic ban differs across the regions (and the provinces) and it depends on many factors such as actual and forecast weather conditions. 
We consider four regions in Northern Italy: Lombardy, Veneto, Piedmont and Emilia Romagna. For each region, the parameter $\lambda$ is estimated averaging the theoretical number of PM$_{10}$ traffic bans that could have happened during the winter semesters $2017-2018$ and $2018-2019$. The data about PM$_{10}$ concentration are provided by \cite{SNPA}. Each day, the regional agency of \cite{SNPA} produces a bulletin indicating the so-called ``day alert'', which happens if the PM$_{10}$ concentration exceeds the safety threshold. 

We calculate the theoretical PM$_{10}$ traffic bans number counting, for each province, the number of times at which the PM$_{10}$ concentration exceeds a threshold of $50$ $\frac{\mu g}{m^3}$ for $4$ consecutive days. Then, for each region, the frequency of PM$_{10}$ traffic bans is calculated averaging the zonal frequencies of traffic bans. For Lombardy region, the annual average traffic bans is $7.2720$. For Veneto region, the annual average number of traffic bans is is $7.4615$.  For Piedmont region, the annual average number of traffic bans is $6.4615$.  For Emilia Romagna region, the annual average number of traffic bans is $9.3300$.\footnote{For Emilia Romagna region, in each zone, the daily PM$_{10}$ concentration is represented by the worst daily concentration. In each zone of the other regions, the daily PM$_{10}$ concentration is calculated averaging the $24$ hours concentration.}

We assume that fossil-fueled vehicle owner's annual discount rate is $5\%$ and that she bears a fixed cost of $150$\euro\ for every traffic ban.\footnote{The cost of each traffic ban is subjective and it represents the economic penalty that the fossil-fueled vehicle owner incurs looking for alternative transportation, such as the inconvenient of different transportation schedule, possible delay and ticket payment.} According to \cite{STATISTA}, Nissan Leaf is the electric vehicle model most sold in the years $2017$ and $2018$ in Italy. Therefore, we set the purchase price, $I$, at $25000$\euro, which is the average purchase cost of Nissan Leaf's vehicle type Acenta. According to the Italian Ministry of Economic Development's Decree of March $20^{\text{th}}$, 2019, when a fossil-fueled vehicle owner decides to scrap a pollutant vehicle and to purchase an electric vehicle whose CO$_2$ emissions are less than $20$ g/km, an incentive of $6000$\euro \ is granted. This is the case of Nissan Leaf Acenta which emits less than $20$ gCO$_2$/km. The parameter values adopter in the analysis are shown in Table ~\ref{table:tabella_parametri}.
\begin{center}
\begin{table}[ht]
\caption{Parameter Description and Estimation.}
\begin{tabular}{lll}
Parameter & Description & Value \\
\hline
$\ell$          &Distance driven   & $12000$    \\
$\frac{a}{b}$  & Mean-reversion level of $X$   & $0.1045$          \\
$b$         & Mean-reversion speed of $X$    & $0.5941$           \\
$\sigma$     & Volatility coefficient of $X$   &  $ 0.090$            \\
$c$          &   Subjective cost associated to one traffic ban  & $150 $   \\
$\lambda_L$          & Frequency of traffic bans in Lombardy case  & $7.2720$   \\
$\lambda_V$          & Frequency of traffic bans in Veneto case  & $7.4615$  \\
$\lambda_P$          &  Frequency of traffic bans  in Piedmont case & $6.4615$   \\
$\lambda_{ER}$          & Frequency of traffic ban in Emilia Romagna case & $9.3300$ \\
$I$       & Electric vehicle purchase cost    & $25000$   \\
$k$       & Incentive offered by policymaker & $6000$        \\
$\rho$         &  Subjective discount rate     &  $0.05$      \\
\end{tabular}
\label{table:tabella_parametri}
\end{table}
\end{center}
\begin{figure}
\includegraphics[scale=0.5]{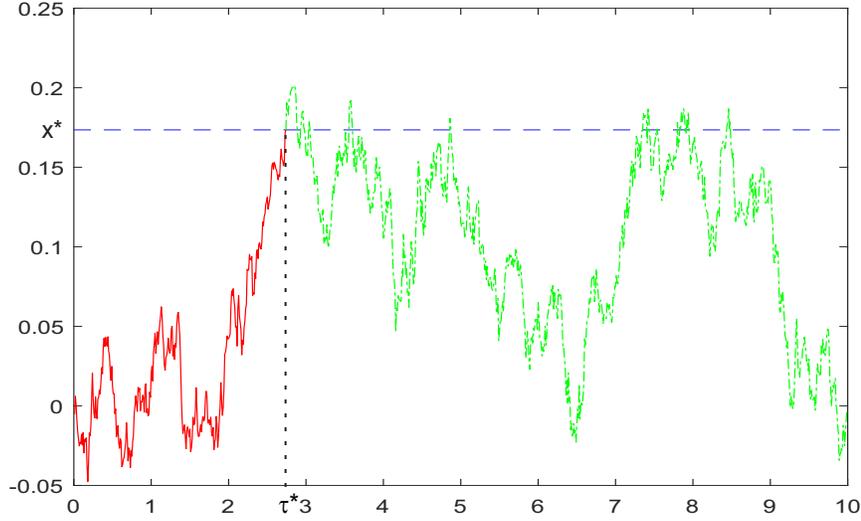}
\caption{An illustration of the optimal switching rule.}
\label{fig:Opt_strategy_graph}
\end{figure}

By using the previous parameters' value, Figure \ref{fig:Opt_strategy_graph} provides an illustrative plot of the optimal switching rule. For the particular scenario presented, a new electric vehicle is bought after $3$ years circa.


\subsection{Comparative Statics}
\label{comparativestatics}

In this section we study both analytically and numerically the sensitivity of the decision threshold $x^{\ast}$ as in \eqref{Statica} with respect to several key parameters. For subsequent considerations, it is important to bear in mind that the optimal switching time is of the form $\tau^*=\inf\{t\geq0:\, X^x_t \geq x^*\}$ so that any monotonicity of $x^*$ with respect to a certain parameter induces the same kind of monotonicity of $\tau^*$.\\
For a given parameter $y \in \left\{ \ell, \lambda, c, k\right\}$, we stress the dependency on $y$ and (with a slight abuse of notation) we write \eqref{Statica} as
$$A^{OU}(x^*(y), y) := \left( x^*(y)  - \beta(y) \right)  \psi_{\rho}^{\prime}(x^*(y))  -\psi_{\rho}(x^*(y)) =0.$$
Then, an application of the implicit-function theorem gives
\begin{equation}
\label{Dini}
\frac{\partial x^{\ast}(y)}{\partial y}=-\frac{A_{y}^{OU}(x^{\ast}(y), y)}{A_{x}^{OU}(x^{\ast}(y), y)}, 
\end{equation}
upon noticing that, because $\psi_{\rho}''>0$ and $x^*(y) > \beta(y)$,
$$
A_{x}^{OU}(x^{\ast}(y), y) = \frac{\partial A^{OU}}{\partial x}(x^*(y), y)= (x^*(y) -\beta(y))\psi^{\prime\prime}_{\rho}(x^*(y)) \ne 0.$$


\subsubsection{Distance traveled}

We start by studying the dependency of the optimal threshold $x^{\ast}$ on the annual distance traveled, $\ell$. By \eqref{Dini} with $y= \ell$ we have
\[
\frac{\partial x^{\ast}(\ell)  }{\partial \ell}=-\frac{A_{\ell}^{OU}(x^{\ast}(\ell), \ell)}{A_{x}^{OU}\left(x^{\ast}(\ell), \ell \right)}.
\]
Because,
\[
\frac{\partial\beta\left(  \ell \right)  }{\partial \ell}=-\frac{\rho+b}{\ell^{2}%
}\left(  I-k-\frac{\lambda c}{\rho}\right)  \text{,}%
\]
we have
\[
A_{\ell}^{OU}\left(x^*(\ell),  \ell \right)  =\psi^{\prime}_{\rho}\left(x^*(\ell)\right)  \left(  \frac{\rho+b}{\ell^{2}}\left(
I-k-\frac{\lambda c}{\rho}\right)  \right)  \text{.}%
\]
Using now that
\[
A_{x}^{OU}\left(x^*(\ell), \ell \right)  = \left(x^*(\ell)-\beta\left(  \ell \right)  \right)  \psi^{\prime\prime}_{\rho}\left(  x^{\ast} (\ell)\right)
\text{,}%
\]
we get from \eqref{Dini}
\[
\frac{\partial x^{\ast}\left(  \ell \right)  }{\partial \ell}=-\frac{\psi^{\prime
}_{\rho}\left(  x^{\ast}\left(\ell \right) \right)  \left(  \frac{\rho+b}{\ell^{2}}\left(  I-k%
-\frac{\lambda c}{\rho}\right)  \right)  }{\left(  x^{\ast}\left( \ell \right) -\beta\left(
 \ell \right)  \right)  \psi^{\prime\prime}_{\rho}\left(  x^{\ast}\left( \ell \right) \right)  }\text{,}%
\]
which, by \eqref{Statica}, also reads as
\[
\frac{\partial x^{\ast}\left(  \ell \right)  }{\partial \ell}= -\frac{\left(  \psi^{\prime
}_{\rho}\left(  x^{\ast}\left( \ell \right) \right)  \right)  ^{2}\left(  \frac{\rho+b}{\ell^{2}}\left(
I-k-\frac{\lambda c}{\rho}\right)  \right)  }{\psi^{\prime\prime
}_{\rho}\left(  x^{\ast} \left( \ell \right) \right)  \psi_{\rho}\left(  x^{\ast} \left( \ell \right) \right)  }.%
\]
We thus have
\begin{equation}
\frac{\partial x^{\ast}\left(  \ell \right)  }{\partial \ell}=\left\{
\begin{array}
[c]{ccc}%
<0 & \text{if} & I-k-\frac{\lambda c}{\rho}>0\\
>0 & \text{if} & I-k-\frac{\lambda c}{\rho}<0
\end{array}
\text{.}\right.
\label{distancecomparata}
\end{equation}

\begin{figure}
 \begin{subfigure}[b]{0.475\textwidth}
            \centering
          \includegraphics[scale=0.5]{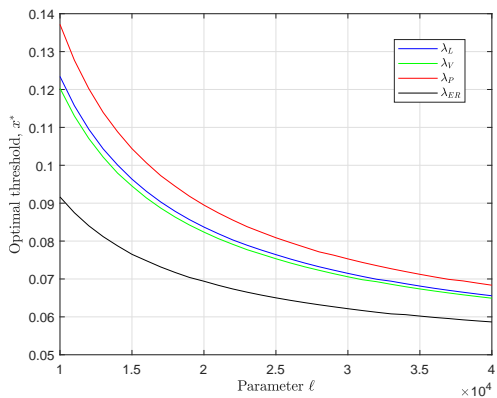}
 \caption[]%
    {{\small Dependency of $x^*$ w.r.t.\ the annual traveled distance, $\ell$, when  $ I-k-\frac{\lambda c}{\rho}>0$.}}
\label{fig:distance_covered_crescente}
        \end{subfigure}
        \hfill
\begin{subfigure}[b]{0.475\textwidth}
\includegraphics[scale=0.5]{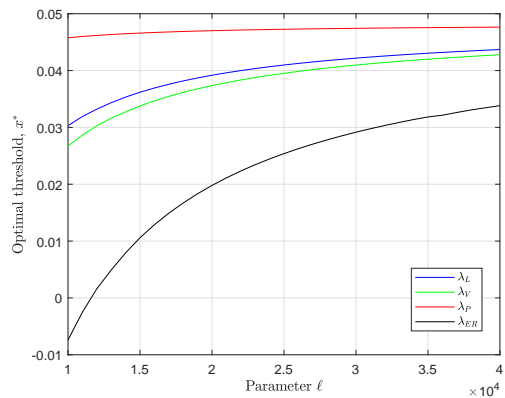}
 \caption[]%
  {{\small Dependency of $x^*$ w.r.t.\ the annual traveled distance, $\ell$, when  $ I-k-\frac{\lambda c}{\rho}<0$.}}
\label{fig:distance_covered_decrescente}
        \end{subfigure}
\caption[St]
        {\small Dependency of $x^*$ w.r.t.\ the annual traveled distance, $\ell$, for different choices of the parameter $\lambda$.}
        \label{}
\end{figure}

A numerical representation of \eqref{distancecomparata} is presented in Figures \ref{fig:distance_covered_crescente} and \ref{fig:distance_covered_decrescente}.

The case $I-k-\frac{\lambda c}{\rho}>0$ is shown in Figure \ref{fig:distance_covered_crescente}. It represents the situation where the purchase cost, $I$, is not completely covered by the policymaker's incentive, $k$, and by the perpetual stream of benefit from avoided traffic bans, $\frac{\lambda c}{\rho}$. Because $k + \frac{\lambda c}{\rho} < I$, the initial investment cost $I$ is more acceptable to the fossil-fueled vehicle owner if the distance traveled increases, as she can then recover the cost $I-k-\frac{\lambda c}{\rho}$ thanks to fuel saving per Kilometer. Under this condition, an increase in the distance traveled leads the fossil-fueled vehicle owner to adopt an electric vehicle sooner. The frequency of traffic bans influences the adoption decision: in the case of Emilia Romagna (black line in Figure \ref{fig:distance_covered_crescente}) characterized by the highest number of traffic bans, the optimal threshold is the lowest (for any distance traveled $\ell$). Conversely, for the red line (Piedmont) where traffic bans are less frequent.

The case $k + \frac{\lambda c}{\rho} > I$ is shown in Figure \ref{fig:distance_covered_decrescente}. It represents the situation in which the purchase cost $I$ is completely covered by the policymaker's incentive, $k$, and by the perpetual stream of benefit from avoided traffic bans, $\frac{\lambda c}{\rho}$. The increasing behavior of the optimal threshold $x^*$ can be explained as follows. The fossil-fueled vehicle owner is subject to the risk that the initial advantage can be significantly eroded by driving costs if electricity costs more than fuel, i.e. if process $X$ becomes negative for significant periods. For the fossil-fueled vehicle owner, whose distance traveled is low, this risk is not perceived as relevant and therefore she agrees to buy an electric vehicle at a lower optimal threshold. Increasing the distance traveled, the risk of negative values of process $X$ starts to be crucial in the purchase decision and she then protects herself by postponing the decision to purchase an electric vehicle and waiting for sufficiently large values of the opportunity cost $X$.


\subsubsection{Frequency of traffic bans}

\begin{figure}
 \begin{subfigure}[t]{0.475\textwidth}
            \centering
 \includegraphics[scale=0.5]{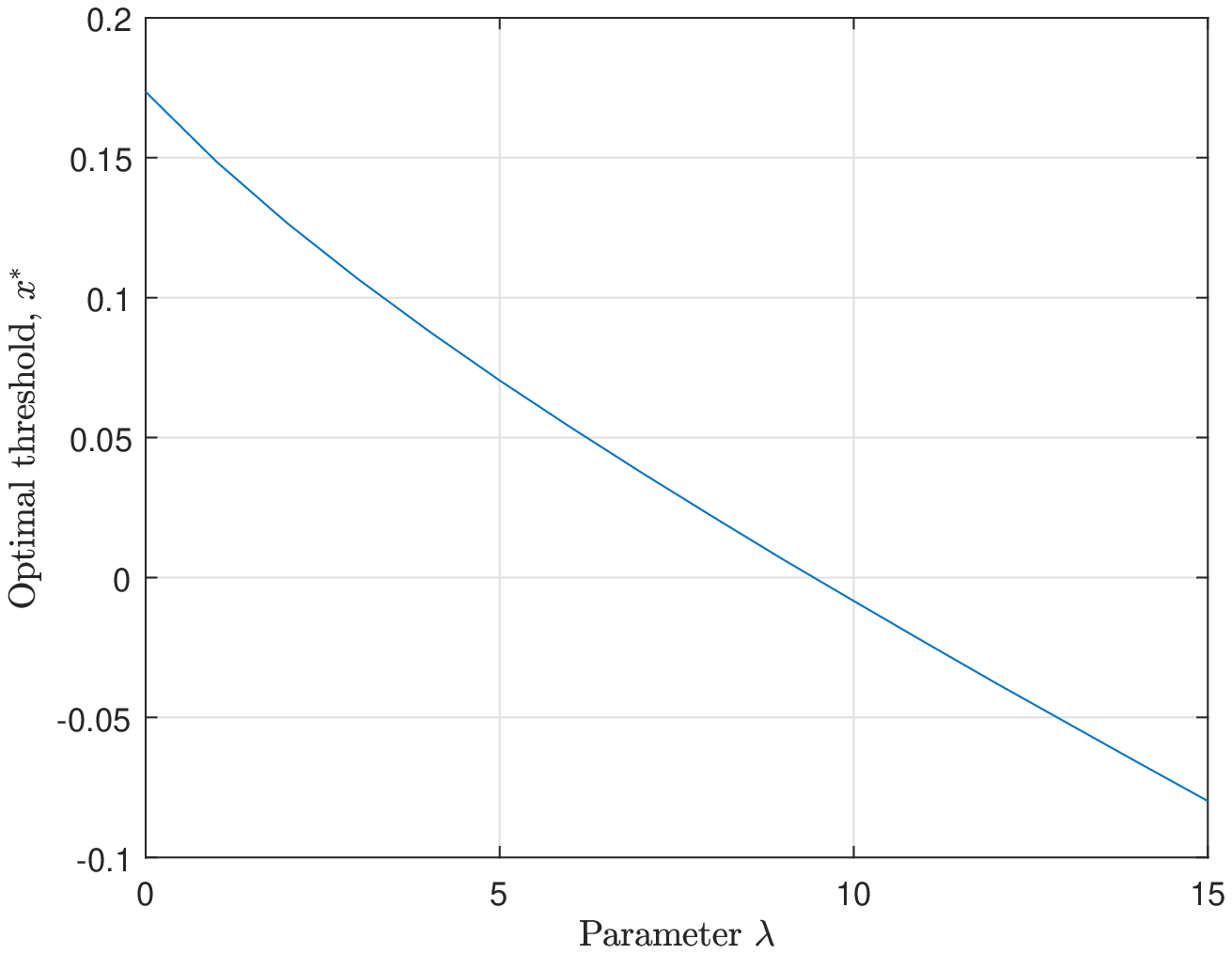}
 \caption[]%
    {{\small Dependency of the optimal threshold $x^*$ w.r.t.\ the frequency of traffic ban, $\lambda$.}}
  \label{PoissonIntensity}
        \end{subfigure}
        \hfill
\begin{subfigure}[t]{0.475\textwidth}
\includegraphics[scale=0.5]{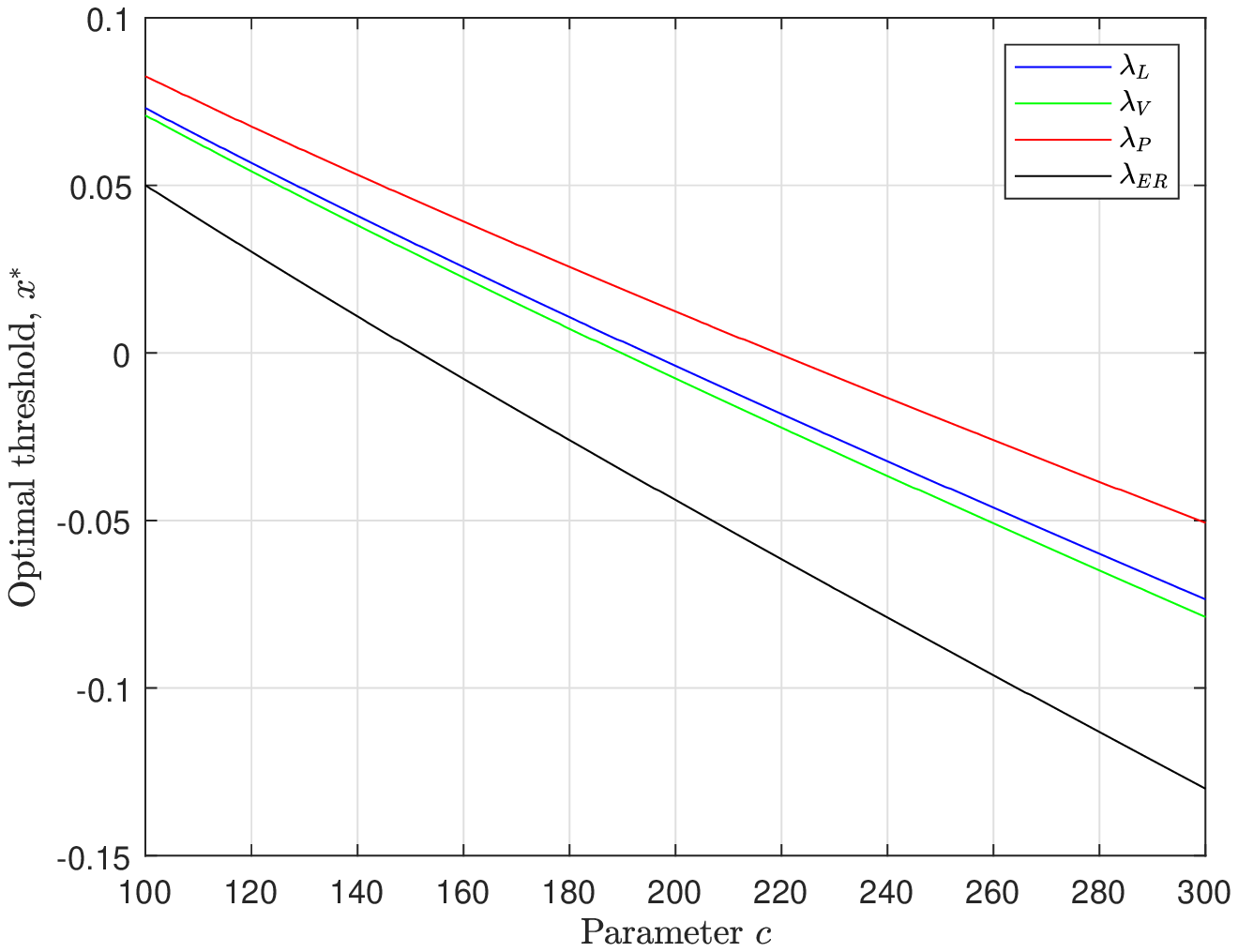}
 \caption[]%
  {{\small Dependency of the optimal threshold $x^*$ w.r.t.\ the subjective cost of one traffic ban, $c$, for different choices of parameter $\lambda$.}}
\label{fig:c_comparata_regioni}
        \end{subfigure}
\caption[St]
        {\small Dependency of $x^*$ w.r.t.\ the frequency of traffic ban, $\lambda$, and the cost of one traffic ban, $c$.}
        \label{}
\end{figure}

We now move on by studying the dependency of the optimal threshold $x^{\ast}$ on the frequency of traffic bans $\lambda$.
Since
\[
A_{\lambda}^{OU}\left(  x^*\left(  \lambda\right)  , \lambda\right)
=\psi^{\prime}_{\rho}\left(  x^*(\lambda)\right)  \cdot\frac{\partial\beta\left(
\lambda\right)  }{\partial\lambda}= \psi^{\prime}_{\rho}\left(  x^*(\lambda)\right)  \left(
\frac{c}{\rho}\frac{\rho+b}{\ell}\right)  \text{,}%
\]
and
\[
A_{x}^{OU}\left(  x^*\left(  \lambda\right)  , \lambda\right)  =\frac{\psi_{\rho}\left(
x^{\ast} \left( \lambda \right) \right)  }{\psi^{\prime}_{\rho}\left(  x^{\ast}  \left( \lambda \right) \right)  }\psi^{\prime\prime
}\left(  x^*(\lambda)\right)  >0\text{.}%
\]
we have, by \eqref{Dini} for $y= \lambda$,
\begin{equation}
\frac{\partial x^{\ast}\left(\lambda\right)  }{\partial\lambda}%
=-\frac{\left(  \psi^{\prime}_{\rho}\left(  x^{\ast} \left( \lambda \right) \right)  \right)  ^{2}\left(
\frac{c}{\rho}\frac{\rho+b}{\ell}\right)  }{\psi_{\rho}\left(  x^{\ast}\left( \lambda \right) \right)
\psi^{\prime\prime}_{\rho}\left(  x^{\ast}\left( \lambda \right) \right)  }<0\text{;}%
\label{trafficbancomparata}
\end{equation}
that is, an increase in the expected number of traffic bans leads to a decrease of the optimal threshold $x^{\ast}$. A numerical representation of \eqref{trafficbancomparata} is shown in Figure \ref{PoissonIntensity}.

If the frequency of traffic bans increases, the optimal threshold $x^*$ decreases and the adoption time decreases as well. 
This can be explained by noticing that an increase in the frequency of traffic bans leads to two main effects: on the one hand, fossil-fueled vehicle owner's driving cost increases; on the other hand, the fossil-fueled vehicle owner cannot circulate during traffic ban periods. As a consequence, she accepts to purchase an electric vehicle at a lower optimal threshold (i.e.\ sooner) because the cost of driving an electric vehicle is then compensated by the savings derived from the avoided traffic bans.


\subsubsection{Cost of one traffic ban}

We here study the dependency of the optimal threshold $x^{\ast}$ on the (subjective) cost associated to one traffic ban, $c$. Because
\[
A_{c}^{OU}\left(  x^* \left(  c\right)  , c\right)  =-\psi^{\prime}_{\rho}\left(  x^* \left( c \right) \right)  \left(  -\frac{\lambda}{\rho}\frac{\rho+b}{\ell}\right) = \psi^{\prime}_{\rho}\left(  x^* \left( c \right) \right)   \frac{\lambda}{\rho}\frac{\rho+b}{\ell}  \text{,}%
\]%
by \eqref{Dini}
\begin{equation}
\frac{\partial x^*\left(  c\right)  }{\partial c}=-\frac{A_{c}^{OU}\left(
x^*\left(  c\right)  , c\right)  }{A_{x}^{OU}\left(  x^*\left( c\right), c\right)  }=
-\frac{\left( \psi^{\prime}_{\rho}\left(  x^* \left( c \right) \right) \right)^2 \left(  \frac{\lambda
}{\rho}\frac{\rho+b}{\ell}\right)  }{\psi_{\rho}\left(  x^{\ast} \left( c \right) \right) \psi^{\prime\prime}\left(  x^* \left( c \right) \right)
}< 0\text{.}%
\label{costbancomparata}
\end{equation}
That is, an increase in the marginal cost of traffic bans leads to a decrease the optimal threshold $x^{\ast}$. A numerical representation of \eqref{costbancomparata} is shown in Figure \ref{fig:c_comparata_regioni}.

The more this cost increases, the earlier the fossil-fueled vehicle owner purchases an electric vehicle. In fact, the disadvantage  associated to lower value of $x^*$ is compensated by the saving derived from the avoided cost of traffic bans that she would incur still having a fossil-fueled vehicle. Of course, in the case of Emilia Romagna (black line in Figure \ref{fig:c_comparata_regioni}) -- which is characterized by a greater frequency of traffic bans -- the optimal threshold is reduced because of the larger traffic bans costs. The opposite holds true for the case of Piedmont (red line), where traffic bans are less frequent.


\subsubsection{Incentive}

\begin{figure}
\includegraphics[scale=0.5]{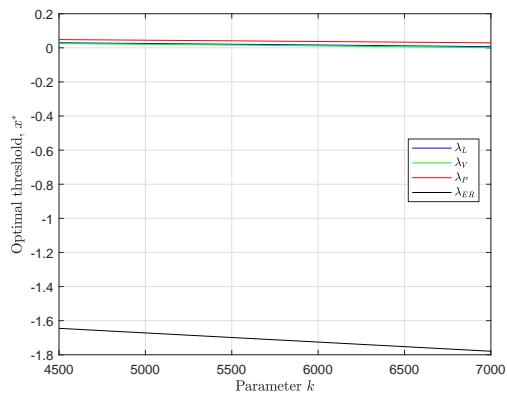}
\caption{Dependency of $x^*$ w.r.t.\ the incentive, $k$, for different choices of the parameter $\lambda$.}
\label{fig:Incentives_comparata}
\end{figure}

We study the dependency of the optimal threshold $x^{\ast}$ on the incentive granted by the policymaker, $k$. Since
$$
A_{k}^{OU}\left(  x^{\ast}\left(  k\right) , k\right)
=\psi^{\prime}_{\rho}(x^*(k))  \left(  \frac{\rho+b}{\ell}\right)  >0,
$$
we find by \eqref{Dini} and $A_{x}^{OU}\left(  x^{\ast}\left(  k\right)  , k\right)= \left(
x^*\left( k \right) -\beta\left(  k \right)  \right)  \psi^{\prime\prime}_{\rho}\left(  x^* \left( k \right) \right) $ that
\begin{equation}
\frac{\partial x^{\ast}\left(  k\right)  }{\partial k}=
-\frac{\left(\psi^{\prime}_{\rho} (x^{\ast}\left(  k\right)   \right)  ^{2}\left(  \frac{\rho+b}%
{\ell}\right)  }{\psi_{\rho} \left( x^{\ast}\left(  k\right)\right)
  \psi^{\prime\prime}_{\rho}\left( x^{\ast}\left(  k\right) \right)  }<0\text{.}%
\label{incentivecomparata}
\end{equation}
An increase in the incentive, $k$, leads to a decrease in the optimal threshold, $x^{\ast}$.  A numerical representation of \eqref{incentivecomparata} is shown in Figure \ref{fig:Incentives_comparata}.

If the policymaker decides to increase the incentive to promote the electric vehicles adoption, the optimal threshold  decreases. That is, the more the incentive increases, the earlier the fossil-fueled vehicle owner is motivated to purchase an electric vehicle. As one can notice traffic bans influence the adoption decision. In Emilia Romagna (black line in Figure \ref{fig:Incentives_comparata}), characterized by a high frequency of traffic bans, the optimal threshold is the lowest one (for any level of incentive) because of the large traffic bans' costs. The opposite happens for the red line (Piedmont), where traffic bans are less frequent.


\subsubsection{Volatility coefficient.} An important question concerns the dependency of the optimal trigger value (hence of the optimal switching time) with respect to the volatility parameter $\sigma$ of the process $X$. Since the fundamental solution $\psi_{\rho}$ also depends on $\sigma$ (cf.\ \eqref{psi-OU}) the analysis of such a sensitivity is slightly more technical than the ones previously performed for $\ell$, $c$, $\lambda$, and $k$. However, arguing as in the proof of Proposition 5.4 in \cite{FK} one can show the next result.

\begin{prop}
\label{comparativa_sigma}
The mapping $\sigma \mapsto x^*(\sigma)$ is increasing.
\end{prop}

\begin{figure*}
\includegraphics[scale=0.5]{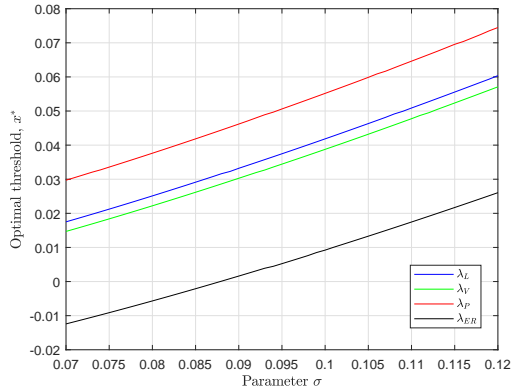}
\caption{Dependency of $x^*$ w.r.t.\ the volatility $\sigma$ for different choices of the parameter $\lambda$.}
\label{fig:sigma_comparata}
\end{figure*}

When the volatility coefficient  $\sigma$ increases, the fossil-fueled vehicle owner is subject to a risk that process $X$ is negative for significant periods, i.e.\ that electricity costs more than fuel.\ Since waiting to invest is a protection against such risk, she best waits for sufficiently large values of the opportunity cost $X$.
A drawing of the conclusion of Proposition \ref{comparativa_sigma} is presented in Figure \ref{fig:sigma_comparata}.

\section{Environmental policy implications: A simulation study}
\label{Opt_time}
In this section, we analyze the expected optimal switching time as a function of the incentive and the frequency of traffic bans. This analysis is particularly relevant for policymakers for many reasons.\ First, it allows to identify a balanced policy mixing incentive and traffic bans to encourage the adoption of an electric vehicle. Moreover, it assesses if such policy reaches the desired impact within a prescribed period of time. \\
In Section \ref{Heuristic discussion and solution}, we proved that the optimal switching time is of the form $\tau^*=\inf\{t\geq0:\, X^x_t \geq x^*\}$.\ 
To conduct the subsequent analysis, we need to stress the dependency of $\tau^{\ast}$ on the frequency rate of traffic bans, $\lambda$, and the incentive, $k$, and, therefore, we write the expected optimal switching time as
\begin{equation}
 \mathbb{E}_x \left[ \tau^{\ast}\left(\lambda,k \right)\right]\text{,}
\label{expected_hitting}
\end{equation}
for any given and fixed $x \in \cI$.\ Observe that $\lambda$ depends on the employed traffic ban rule, so $ \mathbb{E}_x \left[ \tau^{\ast}\left(\lambda,k \right)\right]$ can be seen as an outcome of two distinct policy instruments: the incentive, $k$, and the traffic bans rule.\\
The analysis of \eqref{expected_hitting} is made through Monte Carlo simulation. We set $x=0.02$ and we simulate $10000$ trajectories of an Ornstein-Uhlenbeck process with coefficients $\mu$ and $\sigma$ as specified in Table ~\ref{table:tabella_parametri}.\ For a fossil-fueled vehicle owner, whose features are listed in Table ~\ref{table:tabella_parametri}, we consider a grid of possible combinations of $k$ and $\lambda$. Then, for each possible combination $(\lambda_i,k_i)$, the optimal threshold $x_i^{\ast}:=x^{\ast}\left(\lambda_i,k_i\right)$ is calculated. For each simulated trajectory and for each optimal threshold, the associated optimal switching time, $\tau_i^*:=\tau^{\ast}\left(\lambda_i,k_i\right)$, is then evaluated. Finally, the expected switching time is taken as an average of all the simulated switching times.
\begin{figure}
 \begin{subfigure}[t]{0.475\textwidth}
            \centering
 \includegraphics[scale=0.5]{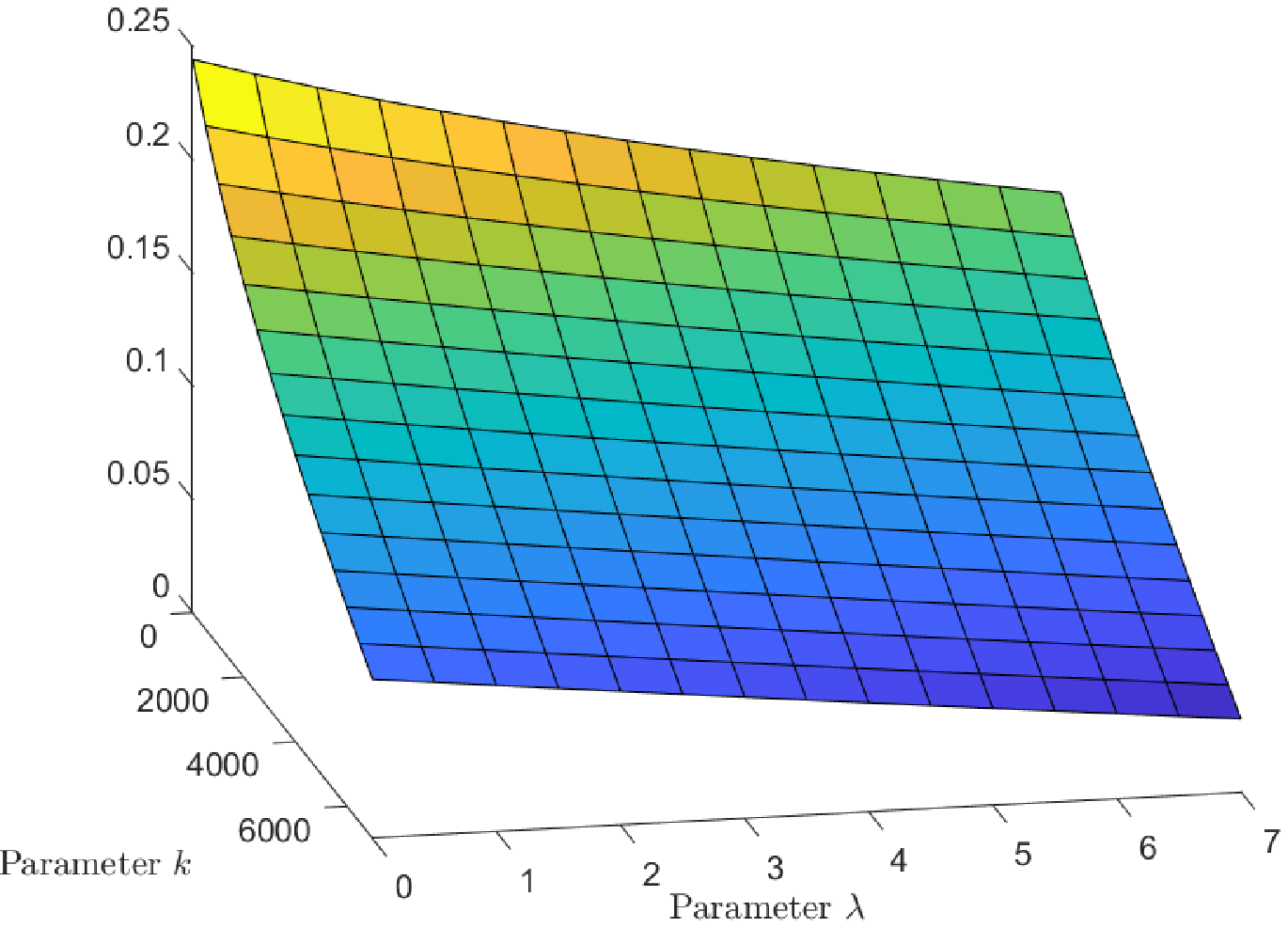}
 \caption[]%
    {{\small Dependency of the expected optimal switching threshold, $x^{\ast}\left(\lambda,k\right)$, w.r.t.\ the frequency of traffic ban, $\lambda$, and the incentive, $k$.}}
  \label{exp_opt_thr}
        \end{subfigure}
        \hfill
\begin{subfigure}[t]{0.475\textwidth}
 \includegraphics[scale=0.5]{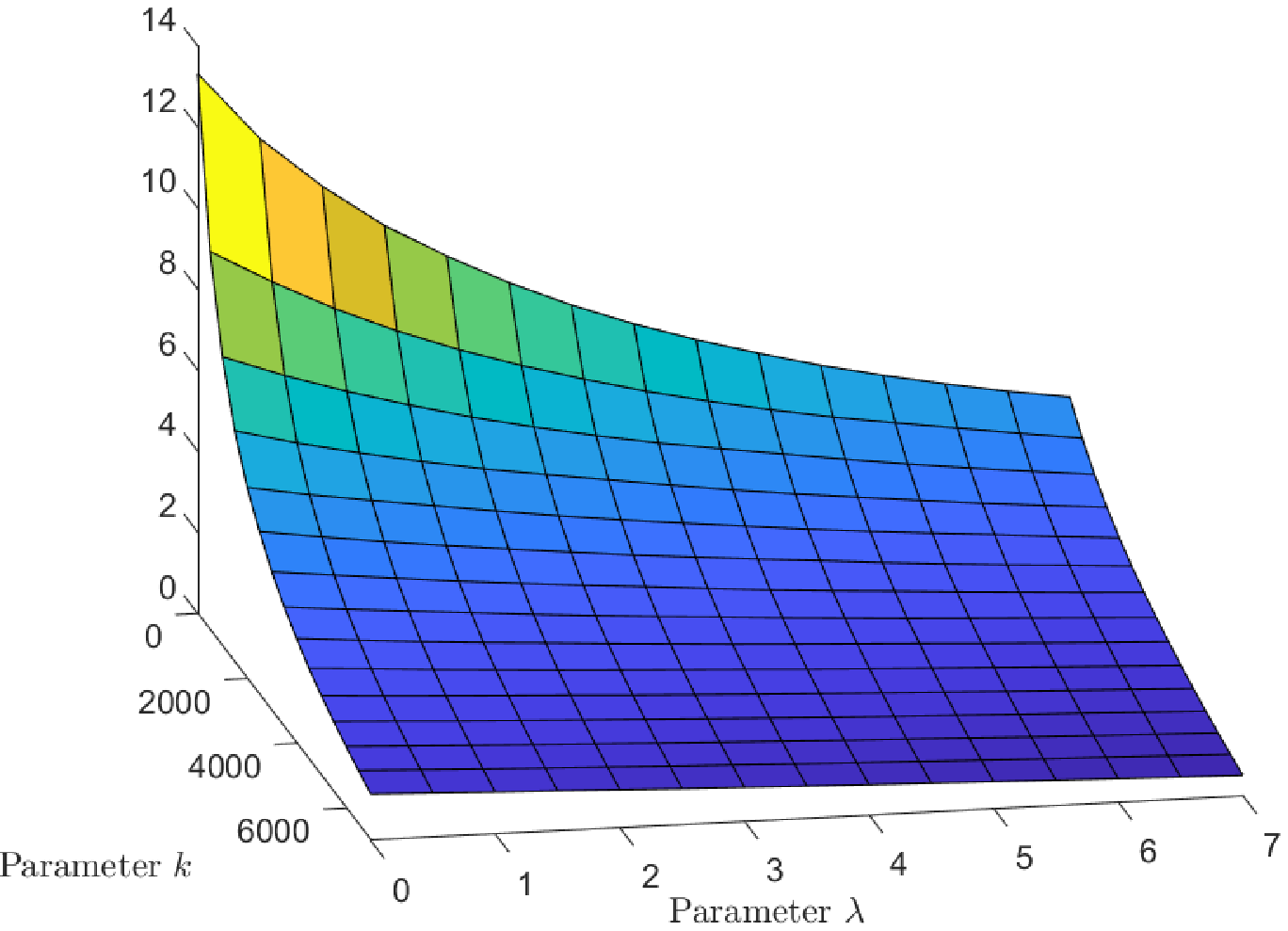}
 \caption[]%
    {{\small Dependency of the expected optimal switching time, $\mathbb{E}_x[\tau^*\left(\lambda,k\right)]$, w.r.t.\ the frequency of traffic ban, $\lambda$, and the incentive, $k$.}}
\label{exp_hit}
        \end{subfigure}
\caption[St]
        {\small Dependency of $x^*\left(\lambda,k\right)$ and $\mathbb{E}_x[\tau^*\left(\lambda,k\right)]$ w.r.t.\ the frequency of traffic ban, $\lambda$, and the incentive, $k$.}
        \label{}
\end{figure}

The switching threshold $x^*$ as a function of $\lambda$ and $k$ is shown in Figure \ref{exp_opt_thr}. The expected optimal switching time as a function of $\lambda$ and $k$ is presented in Figure \ref{exp_hit}. Notice that, as stated in Section \ref{comparativestatics}, any monotonicity of $x^*\left(\lambda,k\right)$ is shared by its associated $\tau^*\left(\lambda,k\right)$, and therefore by  $\mathbb{E}_x[\tau^*\left(\lambda,k\right)]$.\\
Figure \ref{exp_hit}  shows that when both $\lambda$ and $k$ are null, the expected optimal switching time is the highest possible one.\ In particular, the fossil-fueled vehicle owner waits in average circa $13$ years before switching to an electric vehicle. Such a large expected time can be understood by considering, firstly, that she does not incur any traffic ban and, secondly, that the policymaker does not grant any incentive.\ On the opposite case, when both $\lambda$ and $k$ are the greatest possible, the expected optimal switching time is around zero.\\
It is interesting to notice in Figure \ref{exp_hit} that the expected optimal switching time tends to (almost) flat out to a level below $2$ (years) when the frequency of traffic bans exceeds $2 - 4$ per year and the incentive exceeds $3000$\euro.\\
One also observes from Figure \ref{exp_hit} that incentives are more effective than traffic bans. Indeed, in order to see this, it is particularly useful to compare the upper-right and the lower-left corners of the surface in Figure \ref{exp_hit}.\ The upper-right corner represents the effect of the maximum frequency of traffic bans in absence of incentives, while the lower-left corner represents the effect of the maximum incentive in absence of traffic bans. Clearly, the expected optimal switching time at the former is much higher.
\vspace{0.25cm}

\emph{Policymaker implications.} It is relevant to observe that the two policy instruments, traffic bans and incentives, are different. Traffic bans produce temporary reduction of air pollution concentration and create an indirect cost such as, for example, a GDP slowdown.\  However, the key point that encourages the fossil-fueled vehicle owner to purchase an electric vehicle lies in the risk that many possible future traffic bans take place. Indeed, the rules\footnote{With the term ``rule", we mean the number of the consecutive days at which the PM$_{10}$ concentration exceeds the safety threshold of $50$ $\frac{\mu g}{m^3}$.} for imposing a traffic ban are usually adopted without fixing a deadline \footnote{In this paper, traffic bans happen randomly but, in the reality, they are imposed following a given rule, as discusses in Section \ref{numerics}.}. A light traffic ban rule (i.e.\ a traffic containment only after a long period of consecutive high PM$_{10}$ concentration) should lead to a low frequency of traffic bans, and it is therefore proper of a policymaker that has a relatively low environmental concern. On the opposite case, an hard traffic ban rule is used when the policymaker is more interested in improving air quality or when the aim is to prevent the healthcare cost connected to air pollution.\ 
On the other hand, providing public incentives for the adoption of electric vehicles creates a direct cost for the policymaker and, as a consequence, such a policy could have a shorter lifetime with respect to the traffic bans rule. However, as shown in Figure \ref{exp_hit}, incentives appear to be a more effective tool to lead fossil-fueled vehicle owners to switch to an electric vehicle sooner. \\
An environmental policy for the reduction of the air pollution can use traffic bans and incentive separately or jointly. The decision depends, essentially, on the policymaker environmental concern, as well as on the policymaker budget.\ 
The degree of use of those instruments depends, also, on the PM$_{10}$ concentration in the study area.\ A final remark has to be done.  Adopting an environmental policy, the policymaker has to bear in mind that the use of the incentive can be supposed more uncertain, since the decision to accept the incentive or not is entirely in the hands of the fossil-fueled vehicle owner.\ On the other hand, imposing traffic bans is only a decision of policymaker and the fossil-fueled vehicle owner has no other choice but to deal with.
\section{Conclusions}
\label{Conclusion}

In this work, we have provided a real options model for the problem of optimal adoption of an electric vehicle. We consider a fossil-fueled vehicle owner who has to determine the optimal time at which purchase an electric vehicle, while minimizing a cost functional that counts for the distance traveled, the possibility of being randomly stopped for some days, and the net purchase cost. The uncertainty is modeled through the unit distance opportunity cost of driving a fossil-fueled vehicle compared to an electric one. We have solved the resulting optimization problem in the case that the unit distance opportunity cost is modeled as a general one-dimensional It\^o-diffusion and, later, we have provided a model calibration through real data and a detailed comparative statics by assuming a stochastic mean-reversion dynamics. In particular, both in the general case and in the case study, we have completely characterized the critical level of the unit distance opportunity cost triggering the optimal switch to an electric vehicle.

The problem analyzed in this work has provided important consequences both for the policymaker and for the fossil-fueled vehicle owner. A first result concerns the analysis of the parameters that influence the decision of purchasing an electric vehicle: the distance traveled, the subjective cost for each traffic ban, the frequency of traffic bans. With the exception of the driving distance, a change in those parameters implies a monotone response of the adoption time, for any choice of the other model's parameters. On the other hand, the monotonicity of the optimal switching time with respect to the distance traveled is different for different values of the other parameters. In particular, this depends also on how large is the cost of an electric vehicle, net of the incentive received from the policymaker and the saving arising from avoiding traffic bans. 

The second result is that the policymaker has two main tools to lead the fossil-fueled vehicle owner to purchase an electric vehicle: traffic bans rule and incentive. The incentive, from the policymaker point of view, represents a direct cost while imposing a traffic ban involves an indirect cost which is not directly observable, such as, for example, a GDP slowdown. 
Moreover, the policymaker has to balance the two instruments: if the rule to impose traffic bans is tightening, the fossil-fueled vehicle owner is encouraged to switch to an electric vehicle also without granting incentives.\ If the rule to impose traffic bans is light, the policymaker may offer an incentive to encourage the fossil-fueled vehicle owner to switch to an electric vehicle.\ The decision on how combine the two is related to the policymaker's environmental awarness and the policymaker's budget. Future research can focus on the combination of other policies for pollutants reduction, such as, for example, encourage the use of renewable in the electricity production or increasing public transportation.

\appendix
\section{Facts on the underlying diffusion}
\label{AppedixA}

\noindent Here we collect some properties of the process $X$. We refer the reader to Ch.\ II in \cite{BS} for further details. For some reference point $\tilde{x}\in \cI$, we
introduce the derivative of the scale function of $\left\{  X_{t}^{x}\right\}
_{t\geq0}$ as
\begin{equation}
S^{\prime}\left(  x\right)  :=  \exp\left\{  \displaystyle -\int_{\tilde{x}}^x \frac{2\mu\left(
y\right)  }{\sigma^{2}\left(  y\right)  }dy\right\}, \quad  x\in\mathcal{I}\text{.} \label{Sprime}
\end{equation}
Moreover, we introduce the speed
measure density of $\left\{  X_{t}^{x}\right\}  _{t\geq0}$ as
\begin{equation}
m^{\prime}\left(  x\right)  :=\frac{2}{\sigma^{2}\left(  x\right)  S^{\prime
}\left(  x\right)  }, \quad x\in\mathcal{I}\text{.}%
\label{mprime}
\end{equation}

For a given parameter $\rho>0$ (representing in the model the subjective discount factor of the fossil-fueled vehicle owner) we introduce the functions $\psi_{\rho}  $ and $\varphi_{\rho}  $ as the fundamental solutions to the ordinary differential
equation (ODE)
\begin{equation}
\left(  \mathcal{L}_{X}-\rho\right)  u\left(  x\right)  =0 \text{, \ }x\in\mathcal{I}\text{.} \label{ODE_Diff}%
\end{equation}
The function $\psi_{\rho}$ can be chosen to be strictly increasing,
while $\varphi_{\rho} $ strictly
decreasing; both $\psi_{\rho}  $ and $\varphi_{\rho} $ are strictly positive.  The Wronskian between $\psi_{\rho}  $ and $\varphi_{\rho} $ (normalized by the scale function density) is the positive constant
\[
W:=\frac{\psi_{\rho}^{\prime}\left(  x\right)  \varphi_{\rho}\left(  x\right)
-\psi_{\rho}\left(  x\right)  \varphi_{\rho}^{\prime}\left(  x\right)
}{S^{\prime}\left(  x\right)  }> 0 \text{, } \hspace{2mm} x\in\mathcal{I}\text{.}%
\]

For future use, note that, by the linear independence of $\psi_{\rho}  $ and
$\varphi_{\rho}$, any solution to (\ref{ODE_Diff}) can be
written as
\[
u\left(  x\right)  =A\psi_{\rho}\left(  x\right)  +B\varphi_{\rho}\left(
x\right)  \text{,} \hspace{2mm} x \in \cI,
\]
for some suitable parameters $A$ and $B$.

We now recall additional properties of the fundamental solution to (\ref{ODE_Diff}) $\psi_{\rho}$ and $\varphi_{\rho}$.
The fact that $\x$ and $\xx$ are assumed to be natural (i.e.\ unattainable) translate into the analytic conditions:
\begin{equation}
\label{psiphiproperties1}
\lim_{x \downarrow \underline{x}}\psi(x) = 0,\,\,\,\,\lim_{x \downarrow \underline{x}}\varphi(x) = + \infty,\,\,\,\,\lim_{x \uparrow \overline{x}}\psi(x) = + \infty,\,\,\,\,\lim_{x \uparrow \overline{x}}\varphi(x) = 0,
\end{equation}
\begin{equation}
\label{psiphiproperties2}
\lim_{x \downarrow \underline{x}}\frac{\psi'(x)}{S'(x)} = 0,\,\,\,\,\lim_{x \downarrow \underline{x}}\frac{\varphi'(x)}{S'(x)} = -\infty,\,\,\,\,\lim_{x \uparrow \overline{x}}\frac{\psi'(x)}{S'(x)} = + \infty,\,\,\,\,\lim_{x \uparrow \overline{x}}\frac{\varphi'(x)}{S'(x)} = 0.
\end{equation}

Furthermore, for any $\underline{x}<\alpha<\beta<\bar{x}$, one has
\begin{equation}
\frac{\psi_{\rho}^{\prime}\left(  \beta\right)  }{S^{\prime}\left(
\beta\right)  }-\frac{\psi_{\rho}^{\prime}\left(  \alpha\right)  }{S^{\prime
}\left(  \alpha\right)  }=\rho \int_{\alpha}^{\beta}\psi_{\rho}\left(  y\right)
m^{\prime}\left(  y\right)  dy\text{,} \label{Psi_alfa_beta}%
\end{equation}
and
\begin{equation}
\frac{\varphi_{\rho}^{\prime}\left(  \beta\right)  }{S^{\prime}\left(
\beta\right)  }-\frac{\varphi_{\rho}^{\prime}\left(  \alpha\right)
}{S^{\prime}\left(  \alpha\right)  }= \rho \int_{\alpha}^{\beta}\varphi_{\rho
}\left(  y\right)  m^{\prime}\left(  y\right)  dy\text{.}
\label{Phi_alfa_beta}%
\end{equation}

Finally, it is also worth noticing the probabilistic representation of the foundamental solutions $\psi
_{\rho} $ and $\varphi_{\rho} $ in terms of the Laplace transform of hitting times. Letting $\tau_y := \inf\{t \ge 0: X_t =y\}$, $x, y \in \cI$, then
\begin{equation}
\mathbb{E}_{x}\left[  e^{-\rho\tau_y}\right]  =\left\{
\begin{array}
[c]{ccc}%
\frac{\displaystyle \psi_{\rho}\left(  x\right)  }{\displaystyle \psi_{\rho}\left(  y\right)  } &
\text{for} & x<y,\\
\frac{\displaystyle \varphi_{\rho}\left(  x\right)  }{\displaystyle \varphi_{\rho}\left(  y\right)  } &
\text{for} & x>y.
\end{array}
\right.  \label{Laplace}%
\end{equation}

\section{Proof of Theorem \ref{theorem_solution}}
\label{AppendixB}

\begin{proof}
We here prove Theorem \ref{theorem_solution} by following arguments and techniques as those in \cite{Alvarez}, among others.
\vspace{0.25cm}

\emph{Step 1.} We start with the most relevant case in which $\lim_{x \to \x} \left( \ell x + \lambda c \right) < \rho \left( I-k\right) <\lim_{x \to \xx} \left( \ell x + \lambda c \right)$. As already discussed, since $x \mapsto \ell x + \lambda c $ is increasing, we expect that the optimal switching rule is of the form $\tau^* = \inf\{t \ge 0: X_t^x \ge x^*\}$, for some $x^* \in \cI$ to be found. This guess leads to the candidate value function $\widehat{\mathcal{U}}$ given by
\begin{equation}
\widehat{\mathcal{U}}\left(  x\right) :=\left\{
\begin{array}
[c]{lll}%
\left(  I-k-\widehat{V}\left(  x^* \right)  \right)  \mathbb{E}_{x}\left[
e^{-r\tau^{\ast}}\right]  \text{,} & \text{for} & x<x^{\ast} \text{,}\\
I-k-\widehat{V}\left(  x\right)  \text{,} & \text{for} & x\geq x^{\ast} \text{.}
\end{array}
\right.  \label{v_hat_expected_value}%
\end{equation}
Exploiting the probabilistic representation of $\mathbb{E}_{x}\left[
e^{-r\tau^{\ast}}\right]$ as in (\ref{Laplace}), we can write
\begin{equation}
\widehat{\mathcal{U}}\left(  x\right)  =\left\{
\begin{array}
[c]{lll}%
\left(  I-k-\widehat{V}\left(  x^* \right)  \right)  \frac{\psi_{\rho
}\left(  x\right)  }{\psi_{\rho}\left(  x^{\ast}\right)  }\text{,} &
\text{for} & x<x^{\ast} \text{,} \\
I-k-\widehat{V}\left(  x\right)  \text{,} & \text{for} & x\geq x^{\ast} \text{.}%
\end{array}
\right.   \label{v_hat_psi}%
\end{equation}
Notice that $\widehat{\mathcal{U}}$ is already continuous at $x^*$ by construction. In order to determine a candidate for the threshold $x^*$, we impose that $\widehat{\mathcal{U}}$ is $C^1$ at $x=x^*$; i.e. $\widehat{\mathcal{U}}'\left( x^*-\right) =\widehat{\mathcal{U}}'\left( x^*+ \right)$, which in turn leads to
\[
- \left(  I-k-\widehat{V}\left( x^* \right)  \right)  \psi_{\rho
}^{\prime}\left(  x^{\ast}\right) + \left(  I-k-\widehat{V}\right)^{\prime
}\left(  x^{\ast}\right)  \psi_{\rho}\left(  x^{\ast}\right) = 0. 
\label{smoothfit}
\]
Dividing the latter by $S'\left(x \right)$ (cf.\ \eqref{Sprime}) we find
\begin{equation}
\frac{\left(  I-k-\widehat{V}\right)  ^{\prime
}\left(  x^{\ast}\right)  \psi_{\rho}\left(  x^{\ast}\right)  }{S^{\prime
}\left(  x^{\ast}\right)  } - \frac{\left(  I-k-\widehat{V}\left( x^* \right)  \right)  \psi_{\rho
}^{\prime}\left(  x^{\ast}\right)}{S^{\prime} \left(x^* \right) } =0\text{.} \label{24}
\end{equation}
Letting $A: \cI\to \mathbb{R}$ be such that
\begin{equation}
A\left( x \right) := \frac{   \psi_{\rho}\left(  x \right) \left(  I-k-\widehat{V}\right)  ^{\prime
}\left(  x \right)  - \left(  I-k-\widehat{V}\left( x \right)  \right)  \psi_{\rho
}^{\prime}\left(  x \right)}{S^{\prime} \left(x \right) } \label{A}
\end{equation}
we have that \eqref{24} is equivalent to $A\left( x^* \right) =0$.

Using the fact that $S^{\prime}$ solves $\left( \mathcal{L}_X S' \right) \left( x \right) =0$, and the fact that $\psi_{\rho}$ solves \eqref{ODE_Diff}, some algebra shows that
\[
A^{\prime}(x)= \psi_{\rho} \left( x \right) m'\left(x\right) \left( \mathcal{L}_X - \rho \right)\left( I-k - \widehat{V}\right) \left(x \right) \text{.}
\]
Thanks to \eqref{Axbar}, we have that $\lim_{x \to \x} A\left(x \right) =0$; hence, by the foundamental theorem of calculus, for any $x \in \cI$, we have
\begin{equation}
A\left(x \right) = \int_{\x}^x \psi_{\rho} \left(y \right) m'\left( y\right) \left( \mathcal{L}_X - \rho \right)\left( I-k - \widehat{V}\right) \left(y \right) dy \text{.} \label{Ax}
\end{equation}
Since $\left( \mathcal{L}_X - \rho \right) \widehat{V}\left(x \right)=- \left( \ell x + \lambda c \right) $, we can write from \eqref{Ax}
\begin{equation}
A\left(x \right) = \int_{\x}^x \psi_{\rho} \left(y \right) m'\left( y\right) \left(- \rho \left(I-k \right) + \ell y + \lambda c \right) dy \text{,} \hspace{3mm} x \in \cI \text{.} \label{Ax_2}
\end{equation}
Because it must be $A\left( x^* \right)=0$, then we obtain the equation for $x^*$
\begin{equation}
 \int_{\x}^{x^*} \psi_{\rho} \left(y \right) m'\left( y\right) \left( -\rho \left(I-k \right) + \ell y + \lambda c \right) dy =0 \text{.} \label{x_ottimo}
\end{equation}
\vspace{0.25cm}

\emph{Step 2.}  We now show that there exists a unique $x^*$ solving \eqref{x_ottimo} such that $x^* > \hat{x}$ with
$$
\hat{x} := \frac{1}{\ell} \left( \rho \left(I-k\right) - \lambda c \right)\text{.}
$$
With reference to \eqref{Ax_2}, observe that $A\left( \hat{x} \right) <0$ because $ y \mapsto  \left( - \rho \left(I-k\right) + \ell x + \lambda c \right)$ is increasing and null in $\hat{x}$. Using again \eqref{Ax_2} one finds
$$
A^{\prime} \left(x \right) = \psi_{\rho} \left(x \right) m^{\prime} \left(x \right)  \left( - \rho \left(I-k\right) + \ell x + \lambda c \right)
$$
and we observe that $A^{\prime} \left(x \right) >0$ $\forall x > \hat{x}$ and $A^{\prime} \left(x \right) < 0$ $\forall x < \hat{x}$.

Moreover, given $x>\hat{x}+ \delta$, for some $\delta >0$, the integral mean-value theorem and \eqref{Ax} give for some $\xi \in \left( \hat{x}+ \delta, x\right)$
\begin{align*}
& A\left(x \right)  =  \int_{\x}^{x} \psi_{\rho} \left(y \right) m'\left( y\right) \left( - \rho \left(I-k\right) + \ell y + \lambda c \right) dy  \\
&= \int_{\x}^{\hat{x}+\delta} \psi_{\rho} \left(y \right) m'\left( y\right) \left( - \rho \left(I-k \right) + \ell y + \lambda c \right) dy + \int_{\hat{x}+\delta}^{x} \psi_{\rho} \left(y \right) m'\left( y\right) \left( - \rho \left(I-k \right) + \ell y +\lambda c \right) dy \\
& =\int_{\underline{x}}^{\hat{x}+\delta}\psi_{\rho}\left(  y\right)  m^{\prime}\left(
y\right)  \left( - \rho\left(  I-k\right)  + \ell y -\lambda c\right)
dy+\left( \frac{ -\rho\left(  I-k\right)
+ \ell \xi + \lambda c}{\rho}\right) \int_{\hat{x}+\delta}^{x}\rho\psi_{\rho}\left(
y\right)  m^{\prime}\left(  y\right)  dy \\
& =\int_{\underline{x}}^{\hat{x}+\delta}\psi_{\rho}\left(  y\right)  m^{\prime}\left(
y\right)  \left( -  \rho\left(  I-k\right)  +\ell y+ \lambda c\right)
dy+ \left( \frac{ - \rho\left(  I-k\right)
+ \ell \xi + \lambda c}{\rho} \right) \left(  \frac{\psi_{\rho}^{\prime}\left(  x\right)  }{S^{\prime}\left(  x\right)  }-\frac{\psi_{\rho}^{\prime
}\left(  \hat{x}+\delta \right)  }{S^{\prime}\left(  \hat{x} + \delta \right)  }\right)
\text{.}%
\end{align*}
Since $- \rho\left(I-k\right) + \ell \left( \hat{x} + \delta \right)+ \lambda c >0$ and $\lim_{x \uparrow \xx} \frac{\psi_{\rho}^{\prime}\left( x\right)}{S^{\prime} \left( x \right)} = + \infty,$
we have that
$\lim_{x \uparrow \xx} A\left(x \right) = +\infty.$
This fact, together with $A\left( \hat{x}\right)<0$ and $A^{\prime}\left( x \right)>0$ $\forall x > \hat{x}$, leads to the existence of a unique $x^* >\hat{x}$ such that $A\left( x^* \right) =0$; that is, satisfying \eqref{Ax_2}.
\vspace{0.25cm}

\emph{Step 3.} We now prove that the $C^1$-function $\widehat{\mathcal{U}}$ of \eqref{ucapp} is such that
$$
(a)\,\,\, \left(  \mathcal{L}_{X}-\rho\right)  \widehat{\mathcal{U}}\left(  x\right)
=0 \hspace{2mm} \text{on} \hspace{2mm}  x<x^{\ast} \qquad \text{and} \qquad (b) \,\,\, \widehat{\mathcal{U}}\left(  x\right)  =I-k-\widehat{V}\left(  x\right)  \hspace{2mm} \text{on} \hspace{2mm} x\geq x^{\ast},
$$
as well as
\begin{itemize}
\item[(c)] $\widehat{\mathcal{U}}\left(x \right) \le I-k - \widehat{V}\left(x \right)$ \hspace{3mm} $\forall x < x^*$,
\item[(d)] $\left(  \mathcal{L}_{X}-\rho\right)  \widehat{\mathcal{U}}\left(  x\right) \ge 0$ \hspace{7.75mm} $\forall x >x^*$.
\end{itemize}

Since (a) and (b) above are verified by construction, it thus remains to prove (c) and (d).
\vspace{2mm}

\emph{Proof of (c).}  Given \eqref{ucapp} it is enough to show
\begin{equation}
\label{minimum}
\frac{I-k-\widehat{V}\left(  x^*\right)  }{\psi_{\rho}\left(  x^{\ast
}\right)  }\leq\frac{I-k-\widehat{V}\left(  x\right)  }{\psi_{\rho}\left(
x\right)  }\text{,} \hspace{3mm} \forall x < x^*\text{.}
\end{equation}
First of all, we notice that $x^*$ is such that
\begin{equation}
\left.  \frac{d}{dy} \left(  \frac{I-k-\widehat{V}\left(  y\right)  }{\psi_{\rho}\left(
y\right)  }\right)  \right\vert _{y=x^{\ast}%
}= 0
\label{ddy_Ax}
\end{equation}
because
\begin{align*}
\left.  \frac{d}{dy} \left(  \frac{I-k-\widehat{V}  }{\psi_{\rho}\left(
y\right)  }\right)   \right\vert _{y=x^{\ast}%
} & =\left.  \frac{\left(  I-k-\widehat{V}\right)  ^{\prime}\left(  y\right)
\psi_{\rho}\left(  y\right)  -\left(  I-k-\widehat{V}\right)  \left(
y\right)  \psi_{\rho}^{\prime}\left(  y\right)  }{\left(  \psi_{\rho}\left(
y\right)  \right)  ^{2}}\right\vert _{y=x^{\ast}} \\
&=\left. A\left( y\right)  \cdot\frac{W\cdot S^{\prime}\left( y \right)  }{\left(
\psi_{\rho}\left(  y \right)  \right)  ^{2}} \right\vert_{y=x^*}=0\text{,}%
\end{align*}
due to \eqref{ddy_Ax}. Moreover, by \eqref{Ax},
\[
\left.  \frac{d^2}{dy^2} \left(  \frac{I-k-\widehat{V}}{\psi_{\rho}}\right) \left(  y\right)  \right\vert _{y=x^{\ast}}=\left.  \left[ A \left(  y\right)
\cdot\left(  \frac{W\cdot S^{\prime}\left(  y\right)  }{\left(  \psi_{\rho
}\left(  y\right)  \right)  ^{2}}\right)  ^{\prime}+A^{\prime}\left(
y\right)  \cdot\frac{W\cdot S^{\prime}\left(  y\right)  }{\left(  \psi_{\rho
}\left(  y\right)  \right)  ^{2}}\right] \right\vert _{y=x^{\ast}} \text{.}
\]
Now, since $A \left( x^* \right)=0$, using again \eqref{Ax} and the fact that $x^* >\hat{x}$ we have
\begin{align*}
& \left.  \frac{d^2}{dy^2} \left(  \frac{I-k-\widehat{V}}{\psi_{\rho}}\right) \left(  y\right)  \right\vert _{y=x^{\ast}} = A^{\prime}\left(  x^{\ast
}\right)  \cdot\frac{W\cdot S^{\prime}\left(  x^{\ast}\right)  }{\left(
\psi_{\rho}\left(  x^{\ast}\right)  \right)  ^{2}} \\
=& \frac{W\cdot S^{\prime}\left(  x^{\ast}\right)  }{\left(  \psi_{\rho}\left(
x^{\ast}\right)  \right)  ^{2}}\cdot\left.  \frac{d}{dy}\left[  \int%
_{\underline{x}}^{y}\psi_{\rho}\left(  z\right)  m^{\prime}\left(  z\right)
\left( - \rho\left(  I-k\right)  +\ell z + \lambda c \right)  dz\right]
\right\vert _{y=x^{\ast}} \\
& = \psi_{\rho}\left(  x^{\ast}\right)
m^{\prime}\left(  x^{\ast}\right)  \left( -  \rho\left(  I-k\right)
+ \ell x^{\ast} + \lambda c\right)  >0 \text{.}
\end{align*}
This proves that the function $ \frac{I-k-\widehat{V}\left(  x\right)  }{\psi_{\rho}\left(  x\right)  }$ attains a minimum at $x=x^*$ and thus gives \eqref{minimum}.
\vspace{2mm}

\emph{Proof of (d).} For any $x> x^*$ we have
\[
\left(  \mathcal{L}_{X}-\rho\right)  \widehat{\mathcal{U}}\left(  x\right)
=\left(  \mathcal{L}_{X}-\rho\right)  \left(  I-k-\widehat{V}\right)  \left(
x\right)  =-\rho\left(  I-k\right)  +\ell x+\lambda c>0\text{,}%
\]
where in the second equality we use $\left(  \mathcal{L}_{X}-\rho\right)  \widehat{V}\left(  x\right) = - \left( \ell x+\lambda c\right)$ and for the last inequality we use the fact that $x^* > \hat{x}$.
\vspace{0.25cm}

\emph{Step 4.}  The final verification of the optimality of  $\widehat{\mathcal{U}}$ and of $\tau^* =\inf\{ t \ge 0: X_t^x \ge x^*\}$ follows by a standard application of It\^o's formula  (up to a localization argument) and the use of inequalities (a)-(d) above. We refer to \cite{Peskir} for proofs in related settings.
\vspace{0.25cm}

\emph{Step 5.} The proof of the fact that $\widehat{\mathcal{U}}=0$ and $\widehat{\mathcal{U}}=I-k - \widehat{V}$ in the other two cases easily follows by noticing that $V\left( x \right) = \inf_{\tau \ge 0} \mathbb{E}_x \left[ \int_0^{\tau} e^{-\rho t} \left( \ell X_t + \lambda c - \rho \left( I-k \right) \right) dt \right] + \left(I-k \right)$ and \eqref{equazioneValore}.
\end{proof}

\section*{acknowledgements}
The authors thank ARPA Lombardia, ARPA Veneto, ARPA Emilia Romagna e ARPA Piemonte for supplying the data of the day alert. This work was initiated while G.\ Rizzini was visiting the University of Bielefeld, working under the supervision of G.\ Ferrari and M.D.\ Schmeck. G.\ Rizzini thanks the Center for Mathematical Economics (IMW) of Bielefeld University for the hospitality.

\end{document}